\documentclass[11pt]{article}

\usepackage{graphics}
\usepackage{cite}
\usepackage{axodraw}
\usepackage{rotating}
\usepackage{url}
\usepackage{epsfig}
\usepackage{multirow}
\usepackage{float}
\usepackage{placeins}
\usepackage{feynarts}

\makeatletter
\def\paragraph{\@startsection{paragraph}{4}{\z@}{+2.00ex plus
 +1ex minus +.2ex}{1.5ex plus .2ex}{\it\normalsize}}

\expandafter\ifx\csname mathrm\endcsname\relax\def\mathrm#1{{\rm #1}}\fi

\renewcommand{\theequation}{\thesection.\arabic{equation}}
\newcounter{saveeqn}

\@addtoreset{equation}{section}

\makeatother

\def\nl{\nonumber\\}

\def\beq{\begin{equation}}
\def\eeq{\end{equation}}
\def\beqar{\begin{eqnarray}}
\def\eeqar{\end{eqnarray}}
\def\barr#1{\begin{array}{#1}}
\def\earr{\end{array}}
\def\bfi{\begin{figure}}
\def\efi{\end{figure}}
\def\btab{\begin{table}}
\def\etab{\end{table}}
\def\bce{\begin{center}}
\def\ece{\end{center}}

\def\text{\textstyle}

\def\ga{\gamma}

\def\Ga{\Gamma}

\def\refeq#1{\mbox{Eq.~(\ref{#1})}}
\def\refeqs#1{\mbox{Eqs.~(\ref{#1})}}

\def\reffig#1{\mbox{Fig.~\ref{#1}}}
\def\reffigs#1{\mbox{Figs.~\ref{#1}}}

\def\refta#1{\mbox{Tab.~\ref{#1}}}

\def\refse#1{\mbox{Sec.~\ref{#1}}}

\def\citere#1{\mbox{Ref.~\cite{#1}}}
\def\citeres#1{\mbox{Refs.~\cite{#1}}}

\newcommand{\GeV}{\unskip\,\mathrm{GeV}}

\newcommand{\TeV}{\unskip\,\mathrm{TeV}}
\newcommand{\fba}{\unskip\,\mathrm{fb}}

\def\mathswitch#1{\relax\ifmmode#1\else$#1$\fi}
\def\mathswitchr#1{\relax\ifmmode{\mathrm{#1}}\else$\mathrm{#1}$\fi}
\def\mathswitchit#1{\relax\ifmmode{#1}\else$#1$\fi}

\newcommand{\PW}{\mathswitchr W}
\newcommand{\PV}{\mathswitchr V}
\newcommand{\PZ}{\mathswitchr Z}

\newcommand{\Pg}{\mathswitchr g}
\newcommand{\PH}{\mathswitchr H}

\newcommand{\Pe}{\mathswitchr e}

\newcommand{\Pd}{\mathswitchr d}

\newcommand{\Pu}{\mathswitchr u}
\newcommand{\Ps}{\mathswitchr s}
\newcommand{\Pc}{\mathswitchr c}
\newcommand{\Pb}{\mathswitchr b}
\newcommand{\Pt}{\mathswitchr t}
\newcommand{\Pp}{\mathswitchr p}

\newcommand{\Pep}{\mathswitchr {e^+}}
\newcommand{\Pem}{\mathswitchr {e^-}}

\newcommand{\PWp}{\mathswitchr {W^+}}
\newcommand{\PWm}{\mathswitchr {W^-}}

\newcommand{\MV}{\mathswitch {M_\PV}}
\newcommand{\MW}{\mathswitch {M_\PW}}

\newcommand{\MZ}{\mathswitch {M_\PZ}}
\newcommand{\MH}{\mathswitch {M_\PH}}

\newcommand{\Mt}{\mathswitch {m_\Pt}}
\newcommand{\GW}{\mathswitch {\Gamma_\PW}}
\newcommand{\GZ}{\mathswitch {\Gamma_\PZ}}

\newcommand{\scrs}{\scriptscriptstyle}
\newcommand{\sw}{\mathswitch {s_{\scrs\PW}}}
\newcommand{\cw}{\mathswitch {c_{\scrs\PW}}}

\newcommand{\ri}{{\mathrm{i}}}

\newcommand{\rT}{{\mathrm{T}}}

\newcommand{\rd}{{\mathrm{d}}}

\newcommand{\EW}{{\mathrm{EW}}}

\newcommand{\OS}{{\mathrm{OS}}}

\newcommand{\LO}{{\mathrm{LO}}}
\newcommand{\NLO}{{\mathrm{NLO}}}

\newcommand{\DPA}{{\mathrm{DPA}}}

\newcommand{\spp}{\mathswitch {s_{\Pp\Pp}}}

\def\draftdate{\relax}
\def\mda{\relax}
\def\mua{\relax}
\def\mla{\relax}
\def\draft{
\def\thtystars{******************************}
\def\sixtystars{\thtystars\thtystars}
\typeout{}
\typeout{\sixtystars**}
\typeout{* Draft mode!
         For final version remove \protect\draft\space in source file *}
\typeout{\sixtystars**}
\typeout{}
\def\draftdate{\today}
\def\mua{\marginpar[\boldmath\hfil$\uparrow$]%
                   {\boldmath$\uparrow$\hfil}%
                    \typeout{marginpar: $\uparrow$}\ignorespaces}
\def\mda{\marginpar[\boldmath\hfil$\downarrow$]%
                   {\boldmath$\downarrow$\hfil}%
                    \typeout{marginpar: $\downarrow$}\ignorespaces}
\def\mla{\marginpar[\boldmath\hfil$\rightarrow$]%
                   {\boldmath$\leftarrow $\hfil}%
                    \typeout{marginpar: $\leftrightarrow$}\ignorespaces}
\def\Mua{\marginpar[\boldmath\hfil$\Uparrow$]%
                   {\boldmath$\Uparrow$\hfil}%
                    \typeout{marginpar: $\Uparrow$}\ignorespaces}
\def\Mda{\marginpar[\boldmath\hfil$\Downarrow$]%
                   {\boldmath$\Downarrow$\hfil}%
                    \typeout{marginpar: $\Downarrow$}\ignorespaces}
\def\Mla{\marginpar[\boldmath\hfil$\Rightarrow$]%
                   {\boldmath$\Leftarrow $\hfil}%
                    \typeout{marginpar: $\Leftrightarrow$}\ignorespaces}
\overfullrule 5pt
\oddsidemargin -15mm
\marginparwidth 29mm
}

\makeatletter

\def\eqnarray{\stepcounter{equation}\let\@currentlabel=\theequation
\global\@eqnswtrue
\global\@eqcnt\z@\tabskip\@centering\let\\=\@eqncr
$$\halign to \displaywidth\bgroup\hskip\@centering
  $\displaystyle\tabskip\z@{##}$\@eqnsel&\global\@eqcnt\@ne
  \hskip 2\arraycolsep \hfil${##}$\hfil
  &\global\@eqcnt\tw@ \hskip 2\arraycolsep $\displaystyle\tabskip\z@{##}$\hfil
   \tabskip\@centering&\llap{##}\tabskip\z@\cr}
\def\appendix{\par
 \setcounter{section}{0} \setcounter{subsection}{0}
 \def\thesection{\Alph{section}}}

\newcommand{\lsim}
{\;\raisebox{-.3em}{$\stackrel{\displaystyle <}{\sim}$}\;}
\newcommand{\gsim}
{\;\raisebox{-.3em}{$\stackrel{\displaystyle >}{\sim}$}\;}

\def\dsl{\mathpalette\make@slash}
\def\make@slash#1#2{\setbox\z@\hbox{$#1#2$}%
  \hbox to 0pt{\hss$#1/$\hss\kern-\wd0}\box0}

\newcommand{\mr}[1]{{\mathrm{#1}}}

\newcommand{\Pdbar}{{\bar{\Pd}}}

\makeatother

\newcommand{\mvev}{\nu_\mu\mu^+ \Pe^-\bar\nu_\Pe}
\newcommand{\mc}{\mathcal}

\newcommand{\muf}{\mu_\mr{F}}
\newcommand{\mur}{\mu_\mr{R}}

\newcommand{\ed}{\end{document}}

\begin{document}
\thispagestyle{empty}
\def\thefootnote{\fnsymbol{footnote}}
\setcounter{footnote}{1}
\null
\draftdate\hfill FR-PHENO-2016-005, ICCUB-16-021 
\vfill
\begin{center}
{\Large \bf\boldmath
Next-to-leading-order electroweak corrections to 
\\[.5em]
\boldmath{$\Pp\Pp\to\PWp\PWm\to4\,$}leptons at the LHC 
\\[.5em]
\par} \vskip 2em
\vspace{.5cm}
{\large
{\sc B.~Biedermann$^1$, M.~Billoni$^2$, A.~Denner$^1$, S.~Dittmaier$^3$, L.~Hofer$^4$,}
\\[.2cm]
{\sc B.~J\"ager$^2$,  and L.~Salfelder$^2$} } 
\\[.5cm]
$^1$ {\it Julius-Maximilians-Universit\"at W\"urzburg, Institut f\"ur Theoretische Physik und Astrophysik,\\
97074 W\"urzburg, Germany}
\\[0.3cm]
$^2$ {\it Eberhard Karls Universit\"at T\"ubingen, Institut f\"ur Theoretische Physik,\\
72076 T\"ubingen, Germany}
\\[0.3cm]
$^3$ {\it  Albert-Ludwigs-Universit\"at Freiburg, Physikalisches Institut,\\
79104 Freiburg, Germany}
\\[0.3cm]
$^4$ {\it Universitat de Barcelona (UB), Department de F\'isica Qu\`antica i Astrof\'isica (FQA),\\
Institut de Ci\`encies del Cosmos (ICCUB),\\
08028 Barcelona, Spain}
\par 
\end{center}\par
\vskip 2.cm {\bf Abstract:} \par We present results of the first
calculation of next-to-leading-order electroweak corrections to
W-boson pair production at the LHC that fully takes into account
leptonic W-boson decays and off-shell effects.  Employing realistic
event selections, we discuss the corrections in situations that are
typical for the study of W-boson pairs as a signal process
or of Higgs-boson decays \mbox{$\PH\to\PW\PW^*$},
 to which W-boson pair production represents an
irreducible background. In particular, we compare the full off-shell
results, obtained treating the W-boson resonances
in the complex-mass scheme, to previous results in the so-called double-pole
approximation, which is based on an expansion of the loop amplitudes
about the W~resonance poles.  At small and intermediate scales, i.e.\ 
in particular in angular and rapidity distributions, the two
approaches show the expected agreement at the level of fractions of a
percent, but larger differences appear in the TeV range. For
transverse-momentum distributions, the differences can even exceed the
10\% level in the TeV range where ``background diagrams'' with one
instead of two resonant W~bosons gain in importance because of recoil
effects.
\par
\vskip 1.5cm
\noindent
May 2016
\null
\setcounter{page}{0}
\clearpage
\def\thefootnote{\arabic{footnote}}
\setcounter{footnote}{0}


\section{Introduction}
\label{se:intro}

In the 
second phase of data taking at the LHC, which has started in 2015 with a  
centre-of-mass~(CM) energy of $13\TeV$ (almost twice the energy available before)
and a target luminosity of several $100\fba^{-1}$, 
the experimental analyses of many electroweak (EW) processes are
continued both with a deeper 
energy reach into the TeV range and with higher statistics at intermediate energies.
This, in particular, applies to W-boson pair production,
which is not only interesting as signal process
but also as background to
many searches for new physics and to precision studies of the recently discovered
Higgs boson in its $\PW\PW^*$~decay channel.

As signal process, W-pair production is an optimal test-ground for the
triple-gauge-boson interaction of two W~bosons with a photon or a Z~boson,
which is sensitive to physics beyond the Standard Model (SM)
especially at high energies.  Already the experimental analyses of
ATLAS~\cite{ATLAS:2012mec} and CMS~\cite{Chatrchyan:2013yaa} at Run~1
of the LHC provided constraints on non-standard $\gamma\PW\PW$ and
$\PZ\PW\PW$ couplings that are competitive with the results of their
predecessor experiments at the $\Pp\bar\Pp$~collider Tevatron and the
$\Pep\Pem$~collider LEP, but Run~2 of the LHC will further tighten the
existing limits. 
%
%
%
As background, W-pair production is most prominent in the Higgs
production channel $\Pp\Pp\to\PH\to\PW\PW^*$, which did not only play
an essential role in the Higgs-boson discovery, but will provide
important information in precision studies of the coupling of the
Higgs boson to the W~boson.  To this end, both Higgs signal and
irreducible $\PW\PW^*$~background have to be precisely known,
in particular for 
invariant masses below the W-pair production
threshold, where at least one of the W~bosons is off its mass shell.

The above-mentioned phenomenological issues call for theoretical
predictions for W-pair production at the LHC at the highest possible
precision, 
which carefully take into account decay and off-shell effects
of the W~bosons including fully differential kinematics and
phase-space regions below the W-pair threshold.  In order to reach an
accuracy at the few-percent level, radiative corrections of both the
strong and EW interactions have to be calculated and properly
combined.

At leading order (LO), W-pair production at hadron colliders is
dominated by quark--anti\-quark annihilation, $q\bar
q\to\PW\PW\to4\,$fermions.  The largest corrections at
next-to-leading order (NLO) result
from QCD.  They were first calculated for
on-shell W~bosons in \citere{Ohnemus:1991kk}, later refined by 
including leptonic W~decays~\cite{Baur:1995uv}, 
and implemented in
the Monte Carlo program {\tt MCFM}~\cite{Campbell:1999ah}.  As a first
step beyond NLO QCD, the NLO predictions were matched to parton-shower
programs without~\cite{Frixione:2002ik} and including~\cite{Hamilton:2010mb} 
leptonic W~decays. Recently,
complete next-to-next-to-leading order (NNLO) QCD predictions were
presented~\cite{Gehrmann:2014fva} after a continuous effort
over several years~\cite{Chachamis:2008yb}.  Pushing up the inclusive
WW~cross section at the LHC by about $9\%$ ($12\%$) at a CM energy of
$7(14)\TeV$ with respect to NLO QCD, 
the NNLO QCD prediction, which has a residual perturbative uncertainty of $\sim3\%$
in the $q\bar q$ channel, widely relaxed a tension between theory
predictions for inclusive cross sections and LHC data.  In order to
sustain an accuracy of a few 
percent also for fiducial cross sections
based on hard selection cuts, QCD resummations should be taken into
account, which were carried out in
\citere{Grazzini:2005vw}.

Starting at NNLO, W-boson pairs can also be produced in gluon--gluon scattering via quark loops,
$\Pg\Pg\to\PW\PW$, a channel that is enhanced by the large gluon flux at the LHC
and is particularly important in predicting the direct WW background to Higgs production.
At the one-loop level, this channel was considered for on-shell W~bosons in 
\citere{Dicus:1987dj} and later refined by  
including leptonic W~decays~\cite{Binoth:2005ua},
additional jet production~\cite{Melia:2012zg}, $\Pg\Pg\to\PW\PW\Pg\to4\ell\Pg$,
and non-standard couplings~\cite{Bellm:2016cks}.
A comprehensive QCD-based prediction for (off-shell) WW~production at the LHC,
including the $\Pg\Pg$~channel for both WW and WW+jet production,
was presented in 
\citere{Cascioli:2013gfa} in a study of these reactions as background to Higgs-boson production. 
Very recently, the full NLO prediction for $\Pg\Pg\to\PW\PW$ with 
leptonic W-boson decays was presented in the literature~\cite{Caola:2015rqy}. 

At the accuracy level of a few percent, 
EW corrections play an important role as well, in particular due to their known enhancement
by soft/collinear EW gauge-boson exchange at high energies and by
photon emission off
final-state leptons.  Aiming at energies in the
TeV range at the LHC, where EW corrections generically grow to some
$10\%$, in \citeres{Accomando:2004de,Kuhn:2011mh} EW corrections to
W-pair production were first considered in logarithmic approximations
at NLO and NNLO.  Those predictions, however, are valid only in the
Sudakov regime where the absolute values of the Mandelstam
variables $\hat{s},\hat{t},\hat{u}$ of the W~bosons have to be much
larger than the square of the W-boson mass $\MW$. Therefore, the
approximations neither apply to low and intermediate energies, nor to
the dominant 
kinematical domain of forward-produced W~bosons at high
energies.  This shortcoming was overcome by complete NLO EW
calculations for the production of on-shell
W~pairs~\cite{Bierweiler:2012kw,Baglio:2013toa}, which already
revealed large and non-uniform corrections to kinematical
distributions.  A subsequent evaluation of EW corrections that
includes leptonic W~decays and off-shell effects in double-pole
approximation (DPA)~\cite{Billoni:2013aba} confirmed these large EW
corrections and proved their existence in more realistic kinematical
distributions of the W~decay products.  Recently, EW corrections to
leptonically decaying W~pairs were included in {\sc
  Herwig}~\cite{Gieseke:2014gka}, however, in an approach that 
approximately integrates
 out photons
emitted in the WW~production process,
i.e.\ their kinematics is not transferred to the events.

The discussion of EW corrections in the literature also includes
effects of channels with photons in the
initial state (so-called photon-induced channels), 
both for
inelastic~\cite{Bierweiler:2012kw,Baglio:2013toa,Billoni:2013aba,Dyndal:2015hrp}
and elastic~\cite{Luszczak:2013ata} WW~production.  Owing to the huge
$\gamma\gamma\to\PW\PW$ cross section at large partonic CM energies,
the impact of this channel is quite large, in particular in the
forward and backward regions in the TeV range.  On the other hand,
quark--photon channels do not seem to cause 
significant effects because
their impact is swamped by jet radiation from QCD corrections.
Generally, it should be noticed that a precise prediction of the
photon-induced channels is limited by large uncertainties in the
photon distribution function, especially at large momentum fractions
$x>0.1$ of the photon, where uncertainties can be as large as
$100\%$~\cite{Ball:2013hta}. Future fits of parton densities will
certainly improve this situation.

In this paper we further refine the calculations of NLO EW corrections
to W-pair production in $q\bar q$~annihilation by including leptonic
W-boson decays and all off-shell effects of intermediate W~bosons,
comprising, in particular, diagrammatic 
topologies without intermediate
W~pairs (``background diagrams'').  The presented calculation is
thus the first that accounts for EW corrections to the cross section
at partonic CM energies $\sqrt{\hat s}$ below the WW~threshold at
$2\MW$, an issue that, since $\MH<2\MW$, 
is particularly interesting in predicting the
WW~background to Higgs-boson production in the $\PH\to\PW\PW^*$
channel.  Note, however, that the inclusion of
corrections to the background diagrams without intermediate WW~states
is expected to be 
very interesting at high scattering energies as well.  This
can be deduced from the comparison of the full off-shell NLO
prediction for
$\Pep\Pem\to4\,$fermions~\cite{Denner:2005es,Denner:2005fg} with
results in double-pole approximation as delivered by the Monte
Carlo program {\sc RacoonWW}~\cite{Denner:1999kn}.  While the double-pole
approximation is accurate 
within $\sim0.5\%$ for scattering energies
$\lsim500\GeV$, the differences to the full calculation increase for
higher energies.  It will be interesting to perform this comparison
for LHC observables which reach deeply into the TeV range.

In practice, the calculation of NLO EW corrections to
$\Pp\Pp\to4\,$fermions is complex for various reasons.  We have
performed two completely independent evaluations of all ingredients in
order to guarantee reliable and accurate results, 
similar to our NLO calculation for the process
$\Pp\Pp\to\mu^+\mu^-\Pep\Pem+X$~\cite{Biedermann:2016yvs}.
Algebraically, the $2\to4$ one-loop matrix elements are involved, and
an efficient and fast Monte
Carlo integration over the four- and
five-particle phase spaces is a non-trivial task.  One of our
matrix-element calculations follows exactly the strategy of
\citeres{Denner:2005es,Denner:2005fg}, the other one is based on {\sc
  Recola}~\cite{Actis:2012qn}.  To obtain numerically
stable results for the loop integrals everywhere in phase space, we
use the library {\sc Collier}~\cite{Denner:2014gla}.
Conceptually, NLO calculations to processes involving resonances 
are difficult because of issues with gauge invariance, 
since resonance poles are cured by a partial Dyson
summation of self-energy corrections which jeopardizes the validity of
gauge-invariance relations.
We solve this problem by using the so-called complex-mass 
scheme~\cite{Denner:1999gp,Denner:2005fg,Denner:2006ic},
which consistently employs complex gauge-boson masses and delivers
gauge-invariant results at NLO accuracy everywhere in phase phase.

The paper is organized as follows:
In \refse{sec:calculation} we briefly review some details of
our calculation of EW corrections. In particular, we describe the conceptual
difference of the new full off-shell calculation to the previously applied double-pole
approximation. Our phenomenological results are given in \refse{sec:pheno}, comprising
results based on event-selection procedures specifically designed for the investigation
of WW final states and Higgs-boson decays $\PH\to\PW\PW^*$, respectively.
Section~\ref{sec:concl}, finally, contains our conclusions.

%
\section{Details of the calculation}
\label{sec:calculation}

\subsection{Full off-shell calculation}

We consider the proton--proton collision process
\beq
\Pp\Pp \;\to\; \nu_\mu\mu^+ \Pe^-\bar\nu_\Pe+X,
\eeq
which is dominated by the intermediate state of a pair of potentially 
resonant W~bosons decaying leptonically via
$\PWp\to\nu_\mu\mu^+$ and $\PWm\to\Pe^-\bar\nu_\Pe$.
In lowest perturbative order, the process proceeds via the partonic
channels
\beq
\bar qq /  q \bar{q} / \gamma\gamma \;\to\; \nu_\mu\mu^+ \Pe^-\bar\nu_\Pe,
\label{eq:LOchannels}
\eeq
where the contribution of the $\gamma\gamma$ initial state 
is typically strongly suppressed with respect to the dominant
antiquark--quark ($\bar q q/q\bar q$) annihilation channels. 
Figure~\ref{fig:tree-ddnmen} shows the complete set of tree-level diagrams
for the initial state $\Pd\Pdbar$.
\bfi
\centerline{{\small
\unitlength=1.2bp%
\unitlength=1.1bp%








\begin{feynartspicture}(110,180)(1,2)

\FADiagram{}
\FADiagram{}
\FAProp(0.,15.)(4.,10.)(0.,){/Straight}{-1}
\FALabel(0.784783,12.8238)[tr]{$\bar\Pd$}
\FAProp(0.,5.)(4.,10.)(0.,){/Straight}{1}
\FALabel(0.784783,7.17617)[br]{$\Pd$}
\FAProp(20.,20.)(16.5,17.)(0.,){/Straight}{-1}
\FALabel(18.5673,19.1098)[br]{$\ell_1$}
\FAProp(20.,14.)(16.5,17.)(0.,){/Straight}{1}
\FALabel(18.644,14.7931)[tr]{$\ell_2$}
\FAProp(20.,0.)(9.5,10.)(0.,){/Straight}{-1}
\FALabel(15.0576,4.29897)[tr]{$\ell_3$}
\FAProp(20.,7.)(13.,13.5)(0.,){/Straight}{1}
\FALabel(16.8099,9.6468)[tr]{$\ell_4$}
\FAProp(4.,10.)(9.5,10.)(0.,){/Sine}{0}
\FALabel(6.5,9.23)[t]{$\gamma/\PZ$}
\FAProp(16.5,17.)(13.,13.5)(0.,){/Sine}{1}
\FALabel(14.6011,16.1387)[br]{$\PW$}
\FAProp(9.5,10.)(13.,13.5)(0.,){/Straight}{-1}
\FALabel(11.1011,12.6387)[br]{$\ell_3$}
\FAVert(4.,10.){0}
\FAVert(16.5,17.){0}
\FAVert(9.5,10.){0}
\FAVert(13.,13.5){0}

\end{feynartspicture}
\hspace{-4ex}

\begin{feynartspicture}(110,180)(1,2)

\FADiagram{}
\FAProp(0.,15.)(4.,10.)(0.,){/Straight}{-1}
\FALabel(0.784783,12.8238)[tr]{$\bar\Pd$}
\FAProp(0.,5.)(4.,10.)(0.,){/Straight}{1}
\FALabel(0.784783,7.17617)[br]{$\Pd$}
\FAProp(20.,18.5)(14.,15.)(0.,){/Straight}{-1}
\FALabel(17.4139,17.4841)[br]{$\ell_1$}
\FAProp(20.,12.)(14.,15.)(0.,){/Straight}{1}
\FALabel(17.4612,13.0122)[tr]{$\ell_2$}
\FAProp(20.,8.)(14.,5.)(0.,){/Straight}{-1}
\FALabel(17.4798,7.13666)[br]{$\ell_3$}
\FAProp(20.,1.5)(14.,5.)(0.,){/Straight}{1}
\FALabel(17.396,2.81531)[tr]{$\ell_4$}
\FAProp(4.,10.)(9.5,10.)(0.,){/Sine}{0}
\FALabel(6.5,9.23)[t]{$\gamma/\PZ$}
\FAProp(14.,15.)(9.5,10.)(0.,){/Sine}{1}
\FALabel(12.1022,13.1856)[br]{$\PW$}
\FAProp(14.,5.)(9.5,10.)(0.,){/Sine}{-1}
\FALabel(12.2942,6.31276)[tr]{$\PW$}
\FAVert(4.,10.){0}
\FAVert(14.,15.){0}
\FAVert(14.,5.){0}
\FAVert(9.5,10.){0}

\FADiagram{}
\FAProp(0.,15.)(4.,10.)(0.,){/Straight}{-1}
\FALabel(0.784783,12.8238)[tr]{$\bar\Pd$}
\FAProp(0.,5.)(4.,10.)(0.,){/Straight}{1}
\FALabel(0.784783,7.17617)[br]{$\Pd$}
\FAProp(20.,20.)(16.5,17.)(0.,){/Straight}{-1}
\FALabel(18.4173,19.0598)[br]{$\ell_1$}
\FAProp(20.,14.)(16.5,17.)(0.,){/Straight}{1}
\FALabel(18.6402,14.843)[tr]{$\ell_2$}
\FAProp(20.,7.)(13.,13.5)(0.,){/Straight}{-1}
\FALabel(16.8099,9.6468)[tr]{$\ell_3$}
\FAProp(20.,0.)(9.5,10.)(0.,){/Straight}{1}
\FALabel(15.0576,4.29897)[tr]{$\ell_4$}
\FAProp(4.,10.)(9.5,10.)(0.,){/Sine}{0}
\FALabel(6.5,9.23)[t]{$\gamma/\PZ$}
\FAProp(16.5,17.)(13.,13.5)(0.,){/Sine}{1}
\FALabel(14.6011,16.1387)[br]{$\PW$}
\FAProp(13.,13.5)(9.5,10.)(0.,){/Straight}{-1}
\FALabel(11.1011,12.6387)[br]{$\ell_4$}
\FAVert(4.,10.){0}
\FAVert(16.5,17.){0}
\FAVert(13.,13.5){0}
\FAVert(9.5,10.){0}

\end{feynartspicture}
\hspace{-4ex}

\begin{feynartspicture}(110,180)(1,2)

\FADiagram{}
\FAProp(0.,15.)(7.,15.)(0.,){/Straight}{-1}
\FALabel(3.5,16.27)[b]{$\bar\Pd$}
\FAProp(0.,5.)(7.,5.)(0.,){/Straight}{1}
\FALabel(3.5,3.73)[t]{$\Pd$}
\FAProp(20.,18.)(14.,15.)(0.,){/Straight}{-1}
\FALabel(17.3682,17.0936)[br]{$\ell_1$}
\FAProp(20.,12.)(14.,15.)(0.,){/Straight}{1}
\FALabel(17.3682,13.0064)[tr]{$\ell_2$}
\FAProp(20.,8.)(14.,5.)(0.,){/Straight}{-1}
\FALabel(17.4182,7.04361)[br]{$\ell_3$}
\FAProp(20.,2.)(14.,5.)(0.,){/Straight}{1}
\FALabel(17.4182,3.00639)[tr]{$\ell_4$}
\FAProp(7.,15.)(7.,5.)(0.,){/Straight}{-1}
\FALabel(8.07,10.)[l]{$\Pu$}
\FAProp(7.,15.)(14.,15.)(0.,){/Sine}{-1}
\FALabel(10.5,16.07)[b]{$\PW$}
\FAProp(7.,5.)(14.,5.)(0.,){/Sine}{1}
\FALabel(10.5,3.93)[t]{$\PW$}
\FAVert(7.,15.){0}
\FAVert(7.,5.){0}
\FAVert(14.,15.){0}
\FAVert(14.,5.){0}

\FADiagram{}
\FAProp(0.,15.)(4.,10.)(0.,){/Straight}{-1}
\FALabel(0.784783,12.8238)[tr]{$\bar\Pd$}
\FAProp(0.,5.)(4.,10.)(0.,){/Straight}{1}
\FALabel(0.784783,7.17617)[br]{$\Pd$}
\FAProp(20.,20.)(9.5,10.)(0.,){/Straight}{-1}
\FALabel(15.0076,15.801)[br]{$\ell_1$}
\FAProp(20.,13.)(13.5,6.5)(0.,){/Straight}{1}
\FALabel(17.1032,10.4968)[br]{$\ell_2$}
\FAProp(20.,6.)(17.,3.)(0.,){/Straight}{-1}
\FALabel(18.738,5.30504)[br]{$\ell_3$}
\FAProp(20.,0.)(17.,3.)(0.,){/Straight}{1}
\FALabel(18.8032,0.753223)[tr]{$\ell_4$}
\FAProp(4.,10.)(9.5,10.)(0.,){/Sine}{0}
\FALabel(6.5,10.77)[b]{$\gamma/\PZ$}
\FAProp(9.5,10.)(13.5,6.5)(0.,){/Straight}{-1}
\FALabel(11.4928,7.8201)[tr]{$\ell_1$}
\FAProp(13.5,6.5)(17.,3.)(0.,){/Sine}{1}
\FALabel(14.9874,4.35157)[tr]{$\PW$}
\FAVert(4.,10.){0}
\FAVert(9.5,10.){0}
\FAVert(13.5,6.5){0}
\FAVert(17.,3.){0}

\end{feynartspicture}

\hspace{-4ex}

\begin{feynartspicture}(110,180)(1,2)

\FADiagram{}

\FADiagram{}
\FAProp(0.,15.)(4.,10.)(0.,){/Straight}{-1}
\FALabel(0.784783,12.8238)[tr]{$\bar\Pd$}
\FAProp(0.,5.)(4.,10.)(0.,){/Straight}{1}
\FALabel(0.784783,7.17617)[br]{$\Pd$}
\FAProp(20.,13.)(13.5,6.5)(0.,){/Straight}{-1}
\FALabel(16.8182,10.3887)[br]{$\ell_1$}
\FAProp(20.,20.)(9.5,10.)(0.,){/Straight}{1}
\FALabel(14.8076,15.701)[br]{$\ell_2$}
\FAProp(20.,6.)(17.,3.)(0.,){/Straight}{-1}
\FALabel(18.8301,5.19767)[br]{$\ell_3$}
\FAProp(20.,0.)(17.,3.)(0.,){/Straight}{1}
\FALabel(18.2374,1.10157)[tr]{$\ell_4$}
\FAProp(4.,10.)(9.5,10.)(0.,){/Sine}{0}
\FALabel(6.5,10.77)[b]{$\gamma/\PZ$}
\FAProp(13.5,6.5)(9.5,10.)(0.,){/Straight}{-1}
\FALabel(11.383,7.51081)[tr]{$\ell_2$}
\FAProp(13.5,6.5)(17.,3.)(0.,){/Sine}{1}
\FALabel(14.8847,4.04546)[tr]{$\PW$}
\FAVert(4.,10.){0}
\FAVert(13.5,6.5){0}
\FAVert(9.5,10.){0}
\FAVert(17.,3.){0}

\end{feynartspicture}
}
}
\vspace*{-1em}
\caption{Tree-level diagrams for the partonic (charged-current) process
$\bar \Pd\Pd\to4\,$leptons.}
\label{fig:tree-ddnmen}
%
\vspace*{1em}
\centerline{{\small
\unitlength=1.2bp%

\begin{feynartspicture}(110,90)(1,1)

\FADiagram{}
\FAProp(1.,13.)(5.5,13.)(0.,){/Straight}{-1}
\FALabel(0.,12.77)[c]{$\bar\Pd$}
\FAProp(1.,7.)(5.5,7.)(0.,){/Straight}{1}
\FALabel(0.,6.77)[c]{$\Pd$}
\FAProp(18.5,17.)(10.5,17.)(0.,){/Straight}{-1}
\FALabel(20.,16.77)[c]{$\ell_1$}
\FAProp(18.5,13.)(14.5,13.)(0.,){/Straight}{1}
\FALabel(20.,12.77)[c]{$\ell_2$}
\FAProp(18.5,7.)(14.5,7.)(0.,){/Straight}{-1}
\FALabel(20.,6.77)[c]{$\ell_3$}
\FAProp(18.5,3.)(10.5,3.)(0.,){/Straight}{1}
\FALabel(20.,2.77)[c]{$\ell_4$}
\FAProp(5.5,13.)(5.5,7.)(0.,){/Straight}{-1}
\FALabel(6.27,10.)[l]{$\Pu$}
\FAProp(5.5,13.)(10.5,17.)(0.,){/Sine}{-1}
\FALabel(7.97383,15.4152)[br]{$\PW$}
\FAProp(5.5,7.)(10.5,3.)(0.,){/Sine}{1}
\FALabel(7.68279,4.48349)[tr]{$\PW$}
\FAProp(10.5,17.)(14.5,13.)(0.,){/Straight}{-1}
\FALabel(12.1014,14.6014)[tr]{$\ell_2$}
\FAProp(14.5,13.)(14.5,7.)(0.,){/Sine}{0}
\FALabel(15.27,10.)[l]{$\gamma/\PZ$}
\FAProp(14.5,7.)(10.5,3.)(0.,){/Straight}{-1}
\FALabel(12.1032,5.44678)[br]{$\ell_3$}
\FAVert(5.5,13.){0}
\FAVert(5.5,7.){0}
\FAVert(10.5,17.){0}
\FAVert(14.5,13.){0}
\FAVert(14.5,7.){0}
\FAVert(10.5,3.){0}

\end{feynartspicture}
\begin{feynartspicture}(110,90)(1,1)

\FADiagram{}
\FAProp(1.,13.)(5.5,13.)(0.,){/Straight}{-1}
\FALabel(0.,12.77)[c]{$\bar\Pd$}
\FAProp(1.,7.)(5.5,7.)(0.,){/Straight}{1}
\FALabel(0.,6.77)[c]{$\Pd$}
\FAProp(18.5,17.)(10.5,17.)(0.,){/Straight}{-1}
\FALabel(20.,16.77)[c]{$\ell_1$}
\FAProp(18.5,13.)(14.5,13.)(0.,){/Straight}{1}
\FALabel(20.,12.77)[c]{$\ell_2$}
\FAProp(18.5,7.)(14.5,7.)(0.,){/Straight}{-1}
\FALabel(20.,6.77)[c]{$\ell_3$}
\FAProp(18.5,3.)(10.5,3.)(0.,){/Straight}{1}
\FALabel(20.,2.77)[c]{$\ell_4$}
\FAProp(5.5,13.)(5.5,7.)(0.,){/Straight}{-1}
\FALabel(6.27,10.)[l]{$\Pd$}
\FAProp(5.5,13.)(10.5,17.)(0.,){/Sine}{0}
\FALabel(7.97383,15.4152)[br]{$\gamma/\PZ$}
\FAProp(5.5,7.)(10.5,3.)(0.,){/Sine}{0}
\FALabel(7.68279,4.48349)[tr]{$\gamma/\PZ$}
\FAProp(10.5,17.)(14.5,13.)(0.,){/Straight}{-1}
\FALabel(12.1014,14.6014)[tr]{$\ell_1$}
\FAProp(14.5,13.)(14.5,7.)(0.,){/Sine}{1}
\FALabel(15.27,10.)[l]{$\PW$}
\FAProp(14.5,7.)(10.5,3.)(0.,){/Straight}{-1}
\FALabel(12.1032,5.44678)[br]{$\ell_4$}
\FAVert(5.5,13.){0}
\FAVert(5.5,7.){0}
\FAVert(10.5,17.){0}
\FAVert(14.5,13.){0}
\FAVert(14.5,7.){0}
\FAVert(10.5,3.){0}

\end{feynartspicture}
\begin{feynartspicture}(110,90)(1,1)

\FADiagram{}
\FAProp(1.,7.)(5.5,7.)(0.,){/Straight}{-1}
\FALabel(0.,6.77)[c]{$\bar\Pd$}
\FAProp(1.,13.)(5.5,13.)(0.,){/Straight}{1}
\FALabel(0.,12.77)[c]{$\Pd$}
\FAProp(18.5,17.)(10.5,17.)(0.,){/Straight}{-1}
\FALabel(20.,16.77)[c]{$\ell_1$}
\FAProp(18.5,13.)(14.5,13.)(0.,){/Straight}{1}
\FALabel(20.,12.77)[c]{$\ell_2$}
\FAProp(18.5,7.)(14.5,7.)(0.,){/Straight}{-1}
\FALabel(20.,6.77)[c]{$\ell_3$}
\FAProp(18.5,3.)(10.5,3.)(0.,){/Straight}{1}
\FALabel(20.,2.77)[c]{$\ell_4$}
\FAProp(5.5,7.)(5.5,13.)(0.,){/Straight}{-1}
\FALabel(6.47416,10.)[l]{$\Pd$}
\FAProp(5.5,7.)(10.5,3.)(0.,){/Sine}{0}
\FALabel(7.61569,4.39961)[tr]{$\gamma/\PZ$}
\FAProp(5.5,13.)(10.5,17.)(0.,){/Sine}{0}
\FALabel(7.87383,15.5152)[br]{$\gamma/\PZ$}
\FAProp(10.5,17.)(14.5,13.)(0.,){/Straight}{-1}
\FALabel(12.0532,14.7032)[tr]{$\ell_1$}
\FAProp(14.5,13.)(14.5,7.)(0.,){/Sine}{1}
\FALabel(15.4211,10.)[l]{$\PW$}
\FAProp(14.5,7.)(10.5,3.)(0.,){/Straight}{-1}
\FALabel(12.0532,5.34678)[br]{$\ell_4$}
\FAVert(5.5,7.){0}
\FAVert(5.5,13.){0}
\FAVert(10.5,17.){0}
\FAVert(14.5,13.){0}
\FAVert(14.5,7.){0}
\FAVert(10.5,3.){0}

\end{feynartspicture}
}
}
\vspace*{-1em}
\caption{Hexagon diagrams for the partonic (charged-current) process
$\bar \Pd\Pd\to4\,$leptons. The remaining hexagon diagrams are
obtained by reversing the fermion flow in one or both of the fermion
lines of the outgoing fermions and by exchanging
$\ell_1\leftrightarrow \ell_2$ and/or $\ell_3\leftrightarrow \ell_4$.}
\label{fig:hex-ddnmen}
\efi
Analogous diagrams exist for the other down-type quarks, $q=\Ps,\Pb$, and for light up-type quarks, $q=\Pu,\Pc$, while 
top quarks are not considered as active quarks in the proton at LHC energies.
The LO diagrams for $\gamma\gamma$ collisions are, for instance, depicted in
Fig.~1 of \citere{Billoni:2013aba}.

NLO EW corrections, i.e.\ corrections of ${\cal O}(\alpha)$ with
respect to LO, comprise purely EW virtual one-loop diagrams
and real corrections with one additional external photon. 
In total, there are $\mathcal{O}(10^3)$ 
different one-loop diagrams per $\bar qq$ channel.
In \reffig{fig:hex-ddnmen},  we show examples for the
most complicated one-loop topology involving six loop propagators
(so-called hexagon diagrams).

The real photonic corrections are classified into bremsstrahlung corrections
with $\nu_\mu\mu^+ \Pe^-\bar\nu_\Pe+\gamma$ final states, 
\beqar
\bar qq /  q \bar{q} / \gamma\gamma  \;&\to&\; \nu_\mu\mu^+ \Pe^-\bar\nu_\Pe+\gamma,
\eeqar
and photon-induced contributions with an additional $q/\bar q$ in the final state,
\beqar
q\gamma /\gamma q  \;&\to&\; \nu_\mu\mu^+ \Pe^-\bar\nu_\Pe+q,\nl
\bar q\gamma / \gamma \bar q \;&\to&\; \nu_\mu\mu^+ \Pe^-\bar\nu_\Pe+\bar{q}.
\label{eq:qgammachannels}
\eeqar
To simplify our notation in the following, we generically refer to the
$q\bar q / \bar qq$ initial states as $\bar qq$, and to the ones in
\refeq{eq:qgammachannels} as $q\gamma$.  Owing to the suppression of
the LO $\gamma\gamma$ contribution, NLO corrections to $\gamma\gamma$
collisions can be neglected in predictions for $\Pp\Pp$ cross sections
targeting at percent accuracy.  Note that the ${\cal O}(\alpha)$
corrections to $\gamma\gamma\to\PW\PW\to4\,$leptons are known in
DPA~\cite{Bredenstein:2005zk}, without revealing any
unexpected enhancements in this channel.  
We thus base our calculation on the full LO
results including all partonic channels of \refeq{eq:LOchannels}
and on NLO corrections comprising
all virtual EW and real-photonic bremsstrahlung contributions to the
antiquark--quark annihilation channels as well as all the $q\gamma$
contributions of~\refeq{eq:qgammachannels}.

We have performed two independent calculations of all contributions
and find mutual agreement of the squared amplitudes 
at individual phase-space points. 
Cross sections agree within statistical
uncertainties of the final Monte Carlo phase-space integration.

\begin{sloppypar}
The two calculations of the virtual one-loop corrections follow
completely independent 
strategies.
One calculation 
closely follows the approach described in
\citeres{Denner:2005es,Denner:2005fg}, where NLO EW corrections to
$\Pep\Pem\to4\,$fermions via W-boson pairs were calculated,
 and builds on \citere{Billoni:2013aba}
for the real corrections and the Monte Carlo integration.
In detail, the calculation proceeds diagrammatically, starting with the generation
of Feynman diagrams with {\sc FeynArts}~\cite{Kublbeck:1990xc}
and further algebraic processing with in-house {\sc Mathematica} routines.
The other calculation has been carried out with the program
{\sc Recola}~\cite{Actis:2012qn} facilitating the
automated generation of NLO EW amplitudes, 
in combination with an in-house Monte
Carlo generator. %
%
Additional checks have been performed employing the {\sc Mathematica} package {\sc Pole}~\cite{Accomando:2005ra}, which internally makes use 
of {\sc FeynArts}~\cite{Kublbeck:1990xc} and {\sc FormCalc}~\cite{Hahn:1998yk}.
The two loop calculations employ different branches of 
the library {\sc Collier}~\cite{Denner:2014gla},
which is mainly based on the results of \citere{Denner:2002ii},
to evaluate all one-loop integrals with complex W/Z~masses with
sufficient numerical stability in the four-body phase space.
\end{sloppypar}

For the treatment of infrared (soft and/or collinear) singularities, both implementations resort to 
the dipole subtraction approach, as formulated in
\citeres{Dittmaier:1999mb} and \cite{Dittmaier:2008md} for photon radiation
in the cases of collinear-safe and collinear-unsafe
observables, respectively.
Technically, infrared-singular contributions are treated in
dimensional or alternatively in mass regularization, and we 
have checked numerically
that the sum of all (virtual and real) corrections is infrared finite
and independent of the regularization scheme.
Conceptually, we distinguish between leptons that can be fully isolated from
collinear photons and leptons that are recombined with photons in some collinear 
radiation cone. The former case is only relevant for muons and leads to
{\it collinear-unsafe} observables, which receive mass-singular photonic
corrections $\propto(\alpha/\pi)\ln(m_\mu/Q)$, where $Q$ is some hard scale.
On the other hand, the recombination of leptons and collinear photons is
necessary for a 
realistic treatment of electrons which are detected as 
showers in the electromagnetic calorimeter. In this case, the
mass-singular corrections are mitigated to corrections that are
logarithmically sensitive to the resolution parameter of the recombination cone.

Apart from the algebraic complexity, a major complication in the NLO
EW calculation arises from the appearance of resonances which require
at least a partial Dyson summation of self-energy corrections to potentially resonant
propagators, a procedure that jeopardizes the gauge
invariance of the result, if no particular care is taken.  We employ
the complex-mass scheme~\cite{Denner:2005fg,Denner:1999gp,Denner:2006ic} 
which provides
a gauge-invariant solution to this problem at NLO by replacing the
real W- and Z-boson masses by complex quantities, including also the
corresponding complexification of EW couplings.  We emphasize that the
complex-mass scheme maintains NLO precision everywhere in phase space,
i.e.\ in regions with any number of resonant or non-resonant W~bosons.

\subsection{Full off-shell calculation versus double-pole approximation}
\label{sec:4fvsdpa}

Motivated by the dominance of the doubly-resonant contributions to the
cross section of the W-pair production process $\Pp\Pp\to\nu_\mu\mu^+
\Pe^-\bar\nu_\Pe+X$, previous calculations of EW corrections were
based on on-shell
W~bosons~\cite{Bierweiler:2012kw,Baglio:2013toa,Gieseke:2014gka} or an
expansion about the
W~resonances~\cite{Accomando:2004de,Billoni:2013aba}.  More precisely,
the DPA described in \citere{Billoni:2013aba} employed full matrix
elements for all LO contributions and real-photonic corrections and
applied the pole expansion only to the virtual corrections, following
the
approach suggested already for $\Pep\Pem\to\PW\PW\to4f$ at LEP2%
\footnote{Similar DPA variants for W-pair productions at LEP2 were suggested in
\citeres{Beenakker:1998gr,Jadach:1998tz} and compared in \citere{Grunewald:2000ju}.}
and implemented in the Monte
Carlo generator {\sc RacoonWW}~\cite{Denner:1999kn}.

The DPA for the virtual corrections classifies the doubly-resonant loop contributions
into two gauge-invariant categories, known as {\it factorizable} and 
{\it non-factorizable} corrections.
The former comprise all corrections that can be attributed to the production or to the
decays of the resonances; the corresponding generic diagram for this type is shown in
\reffig{fig:factdiag}.
\bfi
\centerline{
\begin{picture}(200,105)(0,0)
\ArrowLine(30,50)( 5, 95)
\ArrowLine( 5, 5)(30, 50)
\Photon(30,50)(150,80){2}{11}
\Photon(30,50)(150,20){2}{11}
\ArrowLine(150,80)(190, 95)
\ArrowLine(190,65)(150,80)
\ArrowLine(190, 5)(150,20)
\ArrowLine(150,20)(190,35)
\GCirc(30,50){10}{.7}
\GCirc(90,65){10}{1}
\GCirc(90,35){10}{1}
\GCirc(150,80){10}{.7}
\GCirc(150,20){10}{.7}
\DashLine( 70,0)( 70,100){2}
\DashLine(110,0)(110,100){2}
\put(50,26){W}
\put(50,68){W}
\put(115,13){W}
\put(115,82){W}
\put(-13, 0){$q$}
\put(-13,95){$\bar q$}
\put(198, 1){$\bar \ell_4$}
\put(198,34){$\ell_3$}
\put(198,60){$\bar \ell_2$}
\put(198,95){$\ell_1$}
\put(-25,-15){\footnotesize On-shell production}
\put(120,-15){\footnotesize On-shell decays}
\end{picture}
}
\vspace*{1em}
\caption{Generic diagram for virtual factorizable corrections to
$\bar qq\to\, \PW\PW\to4\,$leptons appearing in DPA, where the blobs stand for
tree-level or one-loop insertions.}
\label{fig:factdiag}
\efi
%
\bfi
\begin{center}
{\unitlength .8pt \small
\SetScale{.8}
\begin{picture}(360,100)(0,10)
\put(20,0){
\begin{picture}(150,100)(0,0)
\ArrowLine(30,50)( 5, 95)
\ArrowLine( 5, 5)(30, 50)
\Photon(30,50)(90,20){2}{6}
\Photon(30,50)(90,80){-2}{6}
\Vertex(60,65){2.0}
\GCirc(30,50){10}{.0}
\Vertex(90,80){2.0}
\Vertex(90,20){2.0}
\ArrowLine(90,80)(120, 95)
\ArrowLine(120,65)(90,80)
\ArrowLine(120, 5)( 90,20)
\ArrowLine( 90,20)(105,27.5)
\ArrowLine(105,27.5)(120,35)
\Vertex(105,27.5){2.0}
\Photon(60,65)(105,27.5){-2}{5}
\put(86,50){$\gamma$}
\put(63,78){$\PW$}
\put(40,65){$\PW$}
\put(52,18){$\PW$}
\put(-10, 5){$q$}
\put(-10,90){$\bar q$}
\put(128,90){$\ell_1$}
\put(128,65){$\bar \ell_2$}
\put(128,30){$\ell_3$}
\put(128, 5){$\bar \ell_4$}
\end{picture}
}
\put(230,0){
\begin{picture}(120,100)(0,0)
\ArrowLine(30,50)( 5, 95)
\ArrowLine( 5, 5)(30, 50)
\Photon(30,50)(90,80){-2}{6}
\Photon(30,50)(90,20){2}{6}
\GCirc(30,50){10}{.0}
\Vertex(90,80){2.0}
\Vertex(90,20){2.0}
\ArrowLine(90,80)(120, 95)
\ArrowLine(120,65)(105,72.5)
\ArrowLine(105,72.5)(90,80)
\Vertex(105,72.5){2.0}
\ArrowLine(120, 5)( 90,20)
\ArrowLine( 90,20)(105,27.5)
\ArrowLine(105,27.5)(120,35)
\Vertex(105,27.5){2.0}
\Photon(105,27.5)(105,72.5){2}{4.5}
\put(93,47){$\gamma$}
\put(55,73){$\PW$}
\put(55,16){$\PW$}
\put(-10, 5){$q$}
\put(-10,90){$\bar q$}
\put(128,90){$\ell_1$}
\put(128,65){$\bar \ell_2$}
\put(128,30){$\ell_3$}
\put(128, 5){$\bar \ell_4$}
\end{picture}
} 
\end{picture}
} 
\end{center}
\caption{Typical diagrams contributing to the
virtual non-factorizable corrections to
$\bar qq\to\, \PW\PW\to4\,$leptons appearing in DPA, where the blobs stand for
any tree-level subdiagram.}
\label{fig:nonfactdiags}
\efi
The non-factorizable corrections comprise all doubly-resonant contributions
that are due to particle exchange between the various production and decay
subprocesses. Owing to the fact that only soft-photon exchange (as illustrated in
\reffig{fig:nonfactdiags}) leads to doubly-resonant non-factorizable contributions,
they take the form of a single correction factor to the lowest-order amplitude. 
The terminology ``non-factorizable'' refers to the non-trivial off-shell
behaviour of the correction, which is not a simple product of resonance propagators.
More details on the DPA, and more general results on pole approximations, can be found
in \citeres{Denner:1999kn,Beenakker:1998gr,Jadach:1998tz,Accomando:2004de,%
Bredenstein:2005zk,Billoni:2013aba,Dittmaier:2015bfe}.

In order to prepare our comparison of results based on the full
four-fermion ($4f$) calculation to DPA results, we briefly summarize the
most important differences 
between the two approaches:
\begin{itemize}
\item
From the practical point of view, the DPA is simpler to work out, since the 
multiplicities of the underlying loop amplitudes for production and decays 
are much smaller. In our case, the complexity of the DPA loop calculations 
is the one of $2\to2$ production and $1\to2$ decay processes with
about ${\cal O}(10^2)$ diagrams per partonic channel, and 
mostly real particle
masses. On the other hand, the $2\to4$ particle loop calculation involves
${\cal O}(10^3)$ diagrams up to hexagon topology with complex internal masses.

As a result of this difference and due to the possibility of an efficient 
numerical expansion of $2\to2$ loop amplitudes into tree-level-like 
form factors~\cite{Denner:1999kn,Bredenstein:2005zk}, the numerical evaluation of
the DPA can become almost comparable in speed to a
tree-level calculation, 
while the full $4f$ calculation is CPU intensive.
\item
The strength of the full off-shell $4f$~calculation rests in its NLO
accuracy everywhere in phase space, i.e.\ the intrinsic perturbative uncertainty $\Delta_{4f}$
of this approach is generically given by the size of the higher-order corrections
that are not yet calculated. For the purely EW corrections, we thus expect 
$\Delta_{4f}\sim\delta_{\EW}^2$ if $\delta_{\EW}$ is the relative NLO EW
correction factor.

By contrast, the validity of the DPA is restricted to regions in phase
space where the double resonance of the W-boson pair dominates the
cross section.  Taken literally, this restricts the DPA to four-lepton
final states with invariant masses $M_{4\ell}>2\MW+n\Gamma_\PW$ and
$|M_{\ell_i\bar \nu_i}-\MW|\lsim n\Gamma_\PW$, where $n\sim2{-}3$ is
some small number and $\ell_i\bar \nu_i$ generically stands for the two
lepton--neutrino pairs from the W-boson decays.  
To extend the calculation of EW
corrections below the W-pair threshold at 
$M_{4\ell}=2\MW$, an {\it improved
  Born approximation} based on the leading universal corrections is
used for $M_{4\ell}<2 \MW+n\Gamma_\PW$.  In practice, the DPA is
applicable if contributions from
regions below the W-pair threshold and off-shell regions
are sufficiently suppressed.

The theoretical uncertainty $\Delta_{\DPA}$ of the DPA thus is not only given by the
typical size of missing higher-order corrections, but also set by the intrinsic
uncertainty of the pole expansion. Assuming
that all LO contributions are based on full matrix elements and that the relative
correction in DPA, $\delta_{\EW}^{\DPA}$, is normalized to the full LO cross
section $\sigma_{\LO}$, i.e.\
\beq
\sigma_{\NLO\,\EW}^{\DPA} = \sigma_{\LO} + \Delta\sigma_{\EW}^{\DPA}
= \sigma_{\LO}\left(1+\delta_{\EW}^{\DPA}\right), 
\qquad
\delta_{\EW}^{\DPA} = \frac{\Delta\sigma_{\EW}^{\DPA}}{\sigma_{\LO}},
\eeq
we estimate $\Delta_{\DPA}$ to
\beq
\Delta_{\DPA} \;\sim\; \max\biggl\{
\left(\delta_{\EW}^{\DPA}\right)^2, 
\underbrace{\frac{\alpha}{\pi}\frac{\Gamma_\PW}{\MW}\ln(...)}_{\lsim 0.5\%}, 
\left|\delta_{\EW}^{\DPA}\right|\times\frac{|\sigma_{\LO}-\sigma_{\LO}^{\DPA}|}{\sigma_{\LO}^{\DPA}}
\biggr\}.
\label{eq:DeltaDPA}
\eeq
The first term on the r.h.s.\ of \refeq{eq:DeltaDPA} corresponds to the missing higher-order EW corrections,
similar to the NLO limitation of $\Delta_{4f}$. 
The second term indicates the size of the off-shell contributions to the EW corrections 
in regions where the DPA applies. This estimate is based on the typical size of the 
respective effects: the off-shell contributions amounting to a fraction $\sim\Gamma_\PW/\MW$, and the EW corrections being of order $\sim\alpha/\pi$ times some moderate logarithmic factor 
(see also 
\citeres{Denner:1999kn,Denner:2005es,Denner:2005fg,Billoni:2013aba}).
The last term on the r.h.s.\ of \refeq{eq:DeltaDPA} mimics the failure of the DPA upon
blowing up the relative correction $\delta_{\EW}^{\DPA}$ by the factor 
$|\sigma_{\LO}-\sigma_{\LO}^{\DPA}|/\sigma_{\LO}^{\DPA}$ that is deduced from the LO
cross sections based on 
the full $4f$ or DPA matrix elements.%
\footnote{The last uncertainty factor in
  \refeq{eq:DeltaDPA} can also be written as
$\Delta\sigma_{\EW}^{\DPA}/\sigma_{\LO}^{\DPA} \times
|\sigma_{\LO}-\sigma_{\LO}^{\DPA}|/\sigma_{\LO}$, where
$\Delta\sigma_{\EW}^{\DPA}/\sigma_{\LO}^{\DPA}$
%
would then be interpreted as the intrinsic relative EW correction of the DPA
 and $|\sigma_{\LO}-\sigma_{\LO}^{\DPA}|/\sigma_{\LO}$ is the
relative deviation of the DPA LO from the full $4f$ LO cross section.
However, we chose the form of \refeq{eq:DeltaDPA} to be compatible with the definition
of $\delta_{\EW}^{\DPA}$ made in \citere{Billoni:2013aba} and the earlier 
$\Pep\Pem$ references such as \citere{Denner:2005es}.}
As we will see below, the last term in $\Delta_{\DPA}$ is surprisingly large
in some transverse-momentum distributions in the TeV range.
\end{itemize}
%
%
\section{Phenomenological results}
\label{sec:pheno}
\subsection{Input parameters and calculational setup}
\label{ssec:input}
For our numerical studies we consider proton--proton collisions at the LHC at centre-of-mass energies of $\sqrt{\spp} = 8\TeV$
and $13\TeV$. We use the SM input parameters
\beq\label{eq:SMinput}
\begin{array}[b]{rlrrl}
G_\mu &= 1.1663787\times10^{-5}\GeV^{-2}, &\phantom{blablabla}& 
\alpha(0) &=1/137.035999, 
\\
\MW^\OS &=80.385\GeV, &&
\Gamma_\PW^\OS & =2.085\GeV, 
\\
\MZ^\OS &=91.1876\GeV, &&
\Gamma_\PZ^\OS &=2.4952\GeV,
\\
\MH &=125.9\GeV, & &m_\mu&= 0.1057\GeV,
 \\
\Mt&=173.07\GeV, & & \Gamma_\Pt&=2\GeV,
\end{array}
\eeq
following \citere{Beringer:1900zz}.
The muon mass $m_\mu$ is only relevant in the case of a
collinear-unsafe treatment of collinear photon emission off
final-state muons, otherwise 
all fermions except for the top quark
are considered as
massless. 
The finite width of the top quark is only used in
the contributions from $\Pb\gamma$ initial states,
as discussed below.

Throughout, we apply the complex-mass scheme for the W and Z~resonances,
employing complex vector-boson masses $\mu_\PV$ defined by
\beq\label{eq:CMS}
\mu_\PV^2 = M_\PV^2 - \ri M_\PV \Ga_\PV ,  \qquad \PV=\PW,\PZ , 
\eeq
with the real mass values $M_\PV$ and 
constant vector-boson decay widths $\Ga_\PV$.
In this scheme, the weak mixing angle $\theta_\mr{w}$ governing the coupling 
structure of the EW sector is computed from the complex masses,
\beq
\cos\theta_\mr{w}\equiv\cw=\sqrt{1-\sw^2}=\mu_\PW/\mu_\PZ,
\label{eq:cw}
\eeq
and therefore enters the calculation as a complex quantity.
However, the gauge-boson mass and width
values given in \refeq{eq:SMinput} correspond to the
 ``on-shell'' (OS) masses and widths, which were measured at LEP and the
Tevatron using a running-width prescription.
Consequently, we convert these OS
values $M_\PV^{\OS}$ and $\Ga_\PV^{\OS}$ ($\PV=\PW,\PZ$) to the ``pole values'' denoted by $\MV$ and $\Ga_\PV$
according to~\cite{Bardin:1988xt},
\beq\label{eq:m_ga_pole}
\MV = \MV^{\OS}/
\sqrt{1+(\Ga_\PV^{\OS}/\MV^{\OS})^2},
\qquad
\Ga_\PV = \Ga_\PV^{\OS}/
\sqrt{1+(\Ga_\PV^{\OS}/\MV^{\OS})^2},
\eeq
leading to
\beqar
\label{eq:polemasses}
\begin{array}[b]{r@{\,}l@{\qquad}r@{\,}l}
\MW &= 80.357\ldots\GeV, & \GW &= 2.0842\ldots\GeV, \\
\MZ &= 91.153\ldots\GeV,& \GZ &= 2.4942\ldots\GeV.
\label{eq:m_ga_pole_num}
\end{array}
\eeqar
Although the difference between using $\MV^{\OS}$ and $\MV$  would be hardly visible in phenomenological results, we use the latter as input for our numerics.

In the $\bar{q}q$-induced contributions we determine all couplings in the $G_\mu$ scheme, 
where $\alpha$ is defined in terms of the input parameters given in 
\refeqs{eq:SMinput} and (\ref{eq:polemasses}),
\beq
\alpha_{G_\mu} = \frac{\sqrt{2}\,G_\mu \MW^2}{\pi}\left(1-\frac{\MW^2}{\MZ^2}\right).
\eeq
This setting minimizes universal
weak corrections beyond NLO in the high-energy tails of
distributions where high-energy logarithms due to soft/collinear W/Z bosons
dominate the EW corrections.
However, we set $\alpha=\alpha(0)$ for the couplings of the incoming photons in the 
photon-induced ($q\gamma$ and $\gamma\gamma$) channels, since $\alpha(0)$ is the
relevant electromagnetic coupling for real (external) photons.
Therefore, the squared
matrix elements for $q\gamma\to 2\ell2\nu+q$ and $\gamma\gamma\to 2\ell2\nu$
scale like $\alpha(0)\alpha_{G_\mu}^4$ and $\alpha(0)^2\alpha_{G_\mu}^2$,
respectively.

The CKM matrix is set to the unit matrix throughout, without restricting the validity of our
calculation. In fact, since mixing to the third quark generation is phenomenologically negligible
and since we work with mass-degenerate (massless) light quark generations, the CKM
matrix disappears from all amplitudes via its unitarity relations after summing over
intermediate quark states.

\begin{sloppypar}
For the calculation of the hadronic cross section, we use the $\mathcal{O}(\alpha)$-corrected PDF set 
NNPDF2.3QED~\cite{Ball:2013hta}, which also includes a distribution function for the photon.
Since the considered LO cross section and 
the EW corrections do not depend on the strong coupling, all our results exhibit a very weak scale dependence.
We therefore see no benefit in introducing a specific dynamical scale, 
but choose a fixed factorization and renormalization scale setting
\beq
\label{eq:scalechoice}
\muf = \mur = \MW,
\eeq
as the default. Following the arguments of \citere{Diener:2005me}, we employ
a DIS-like factorization scheme for 
the QED corrections because
EW corrections are not taken into account in the fit of the PDFs to data.
\end{sloppypar}

In subprocesses with a final-state quark or antiquark, a parton $i$ is only considered as jet
if its transverse momentum $p_{\rT,i}$ and its rapidity $|y_i|$ allow for a proper detection in experiment. For a parton $i$ we require 
\beq
\label{eq:jet-def}
p_{\rT,i} > p_\mr{T,jet}^\mr{def} = 25\GeV, \qquad |y_i|< |y_{\mr{jet}}^\mr{def}|=5,
\eeq
for being treated as a jet. If the parton does not meet these two requirements, it is treated as
{\it invisible jet}, which means that no additional jet-related cuts are applied and its momentum contributes to $\vec{p}_\rT^\mr{\;miss}$.
A similar treatment is applied to 
final-state photons with rapidities outside the range accessible to the detector, which therefore remain undetected.
We consider final-state photons with
\beq
|y_\gamma| > 5,
\eeq
as {\it invisible photons}, and the corresponding momentum is part of the missing momentum.

As mentioned above, we provide two different setups concerning the treatment of nearly collinear photons:
In the {\it collinear-safe setup}, nearly collinear photons are recombined with 
both final-state electrons and muons, whereas
in the {\it collinear-unsafe setup}, we apply the photon recombination procedure only to final-state electrons
while photons are never recombined with muons.
The applied recombination procedure mimics the experimental concept of ``dressed leptons'' used by ATLAS 
(see e.g.\ \citere{Aad:2011gj}),
which avoids 
the experimental problem to separately resolve an electron and a nearly collinear photon.
As measure for the collinearity of photons and leptons,
their distance $R_{ij}$ in the rapidity--azimuthal-angle plane is used, 
where
\beq
R_{ij} = 
\sqrt{
(y_i-y_j)^2+\Delta\phi_{ij}^2
}
\eeq
with $y_{i,j}$ denoting the rapidities of particle $i$ and $j$ and $\Delta\phi_{ij}$ their azimuthal angle difference,
respectively.
Whenever the separation is smaller than
\beq
R_{\gamma \ell}^\mr{recomb}=0.1,
\eeq
we add the photon momentum to the respective lepton and discard the photon from the event,
while the momenta of all other particles in the event remain unaffected.

\subsection{Event selection}
\label{ssec:cuts}

We compare results for three different event-selection setups:
(i) defined by basic particle identification only (``inclusive setup''),
(ii) designed by ATLAS for enhancing the WW signal (``ATLAS WW setup''), and
(iii) inspired by Higgs-boson analyses in the decay channel $\PH\to\PW\PW^*$
(``Higgs-background setup''), where direct W-pair production appears as irreducible background.
For the latter two setups we present results for integrated and differential cross sections,
for the inclusive setup we show integrated results only.

\subsubsection*{(i)~Inclusive setup}
After a potential 
photon recombination we define the events for the process $\Pp\Pp\to \mvev +X$ by requiring 
the $\mu^ +$ and $\Pe^-$ to have 
transverse momenta
\beq
\label{eq:ptl-cut}
p_{\rT,\ell} > 20\GeV,
\eeq
in the central rapidity region of the detector,
\beq
\label{eq:yl-cut}
|y_\ell| < 2.5.
\eeq
For final states with an identified jet (cf.~\refeq{eq:jet-def}) we demand this jet to be well separated from
the lepton system, by rejecting any event with
\beq
R_\mr{jet,\ell}<0.4.
\eeq 
%
In order to suppress overwhelmingly large QCD corrections from additional
jet radiation, we employ a jet veto, i.e.\ we reject any event with
\beq
\label{eq:incljet-veto}
p_\mr{T,jet}>100\GeV.
\eeq

\subsubsection*{(ii) ATLAS WW setup}

Moreover, we consider an event selection that is inspired by the analyses of the $7\TeV$
dataset for $\Pp\Pp\to \PWp\PWm+X$ performed by the ATLAS collaboration~\cite{ATLAS:2012mec}.
In addition to the afore-mentioned lepton cuts we impose a stronger transverse-momentum cut on the hardest charged lepton and require the two charged leptons to be well separated from 
each other by imposing the cuts 
\beqar
\label{eq:ATLAScut1}
p_{\rT,\ell}^\mr{leading} > 25\GeV,
\qquad
R_{\Pem\mu^+}>0.1,
\qquad
M_{\Pem\mu^+} > 10\GeV.
\eeqar
For a cleaner signature we further demand a non-vanishing missing transverse momentum,
\beqar
\label{eq:ATLAScut2}
E_\rT^\mr{miss} = |\vec{p}_\rT^\mr{\;miss}| > 25\GeV,
\eeqar
and remind the reader that invisible jets and invisible photons do also 
enter this quantity.
To further suppress the influence of QCD corrections, we veto all events with hard final-state jets 
obeying
\beq
\label{eq:ATLASjet-veto}
p_{\rT,\mr{jet}}>25\GeV.
\eeq
Note that according to \refeq{eq:jet-def}, 
all events with a detected jet 
are discarded due to this cut.

\subsubsection*{(iii)~Higgs-background setup}

Inspired by the recent analyses of the decay of the Higgs boson to $\PW\PW^*$ independently performed by ATLAS~\cite{ATLAS:2014aga} and CMS~\cite{Chatrchyan:2013iaa}
 we study the influence of EW corrections on the main irreducible background, namely $\Pp\Pp\to \PW\PW^*\to \mvev +X$, in a realistic cut scenario.
Essentially following \citere{ATLAS:2014aga}, we extend the ATLAS~WW setup of 
Eqs.~(\ref{eq:ATLAScut1})--(\ref{eq:ATLASjet-veto}) by two additional cuts,
\beq
\label{eq:HtoWWcut}
10\GeV < M_{\Pem\mu^+} < 55\GeV,\qquad \Delta\phi_{\Pem\mu^+}<1.8,
\eeq
which are designed to favour the signal topology of the H$\to\PW\PW^*$ analysis and 
therefore significantly suppress direct $\PW\PW^*$ production.
Additionally, we adjust the threshold of the transverse-momentum cut
to the value used in the experimental analysis,
\beq
\label{eq:HtoWWcut2}
E_\rT^\mr{miss} = |\vec{p}_\rT^\mr{\;miss}| > 20\GeV.
\eeq

\subsection{Results on integrated cross sections}

In \refta{tab:xsecs_extended} we present the quark-initiated LO cross
sections, $\sigma^\mr{LO}_{\bar q q}$, for $\Pp\Pp\to\mvev+X$ at the
LHC for the energies $\sqrt{\spp}=8\TeV$ and $\sqrt{\spp}=13\TeV$
within the three different setups defined in \refse{ssec:cuts}.
%
\begin{table}
\begin{center}
\begin{tabular}{|cc|c|ccc|c|c|}
\hline
\multicolumn{2}{|c|}{LHC}&$\sigma^\mr{LO}_{\bar q q}$~[fb] &  $\delta^\mr{NLO}_{\bar q q}$~[$\%$]& $\delta_{q \gamma}^{q\neq\Pb}$~[$\%$] & $\delta_{\gamma\gamma}$~[$\%$] & $\delta_{\mr {EW}}$~[$\%$]& $\delta_{\Pb \gamma}$~[$\%$] 
\\
\hline
\hline
\multirow{2}{*}{Inclusive} & $8\TeV$ &238.65(3) & $-3.28$ & $0.44$   & $0.84$ & $- 2.01$ & $1.81$ \\ 
&$13\TeV$ &390.59(3)& $-3.41$  & $0.49 $  &$0.73$  &  $-2.20$& $2.30$ \\
\hline
\hline
\multirow{2}{*}{ATLAS~WW} & $8\TeV$ &165.24(1) & $-3.56$ & $-0.26$   & $1.01$  &$- 2.81$ &$0.18$ \\ 
&$13\TeV$ &271.63(1)& $-3.71$ & $-0.27$  &$0.87$   &$-3.11$ & $0.23$ \\
\hline
\hline
\multirow{2}{*}{Higgs~bkg} & $8\TeV$ &  $31.59(2)$ & $-2.52$ & $-0.21$   & $0.60$   & $-2.13$&$0.15$ \\ 
&$13\TeV$ &$49.934(2)$  & $-2.54$  &$-0.22$  &$0.52$  & $-2.25$&$0.18$ \\
\hline
\end{tabular}
\caption{
\label{tab:xsecs_extended}
LO cross sections and relative EW corrections
to $\Pp\Pp\to \mvev+X$ at the LHC at $8\TeV$ and $13\TeV$, 
in the inclusive setup (top), the ATLAS~WW setup (middle), and the Higgs-background setup (bottom).
The numbers in parentheses represent the Monte Carlo uncertainty on the last given digit.
}
\end{center}
\end{table}
The numbers include the five $\bar qq$-initiated contributions
induced by the light quarks.
Due to the small bottom-quark PDFs, the $\bar\Pb\Pb$~channel 
comprises less than $2\%$ of the LO cross section in all considered setups.
The LO $\gamma\gamma$-induced subprocess gives rise to a (sub-)percent correction 
to $\sigma^\mr{LO}_{\bar qq}$, so that we treat 
$\delta_{\gamma\gamma}=\sigma^\mr{LO}_{\gamma\gamma}/\sigma^\mr{LO}_{\bar qq}$
as part of the EW corrections.

The NLO corrections of $\mathcal{O}(\alpha)$ consist of virtual one-loop and
photonic real-emission contributions to the $\bar q q$-induced
processes, as well as additional $q\gamma$-,  $\Pb\gamma$-, and
$\gamma\gamma$-initiated tree-level subprocesses.
Similarly to the situation at LO, the corrections stemming from
subprocesses with initial-state bottom-quark pairs account for less
than $2\%$ of the full NLO correction.  For this reason it is
justified to approximate
the matrix elements with 
 external bottom quarks in the $\bar qq$ contributions
by those with
 massless down quarks
 (and internal massless up-type weak isospin partner)
in the EW corrections, without any notable     
loss in precision.
We follow this approach in our diagrammatic loop
calculation, while the second calculation with {\sc Recola} does not
make use of this approximation.%
\footnote{In all considered setups, we find the
  difference on $\delta^\mr{NLO}_{\bar q q}$ to be
  below $0.02\%$ 
  between the two approaches.
  The numbers in \refta{tab:xsecs_extended} are obtained in the approximation where the top quark is treated as massless. 
  }
As default setup for the photonic real-emission contributions we choose the collinear-unsafe setup, 
as defined in \refse{ssec:input}. 
The difference to the collinear-safe setup is discussed in detail in \refse{ssec:safevsunsafe} below.

The omission of virtual and photonic
real-emission corrections of $\mathcal{O}(\alpha)$ to the
$\gamma\gamma$-induced contributions is justified by their small size
already at LO.
While we include the $\bar \Pb\Pb$-induced processes in $\bar q q$, we
display the $\Pb\gamma$-initiated contributions separately from
$q\gamma$ (containing $q=\Pu,\Pd,\Ps,\Pc$) 
as these are the only contributions with bottom quarks in the final state.
Note further that the $\Pb\gamma$-induced contributions are enhanced
by the presence of a top resonance, in particular for the inclusive
setup where they reach $2.3\%$.
In the ATLAS WW setup and the
Higgs-background setup the contributions of this resonance are
suppressed by the jet veto of \refeq{eq:ATLASjet-veto}.  In the
following results on differential distributions for the latter two setups 
we do not include these contributions, which could be further suppressed by a $\Pb$-jet
veto, i.e.\ we identify
$\delta_{q \gamma} \equiv \delta_{q \gamma}^{q\neq\Pb}$.

In \refta{tab:xsecs_extended}, besides the LO cross sections, we list
the relative contributions of the different types of corrections
normalized to $\sigma^{\mr{LO}}_{\bar q q}$.  The generally negative
$\mathcal{O}(\alpha)$ corrections ($\delta^\mr{NLO}_{\bar{q}q}$ and
$\delta_{q\gamma}$) are somewhat reduced by the positive LO
$\gamma\gamma$ correction, so that their sum ($\delta_\mr{EW}$) leads
to a small negative correction of roughly $-(2{-}3)\%$ on the
integrated cross section for all investigated setups.  
However,
as well known from previous studies of W-boson pair production at
hadron
colliders~\cite{Accomando:2004de,Kuhn:2011mh,Bierweiler:2012kw,Baglio:2013toa,Billoni:2013aba},
the EW corrections become very important in differential distributions
where they lead to significant distortions.

\subsection{Results on differential cross sections in the ATLAS WW setup}

In this section we inspect some important differential cross sections
evaluated in the ATLAS WW setup for the collinear-unsafe
photon scenario. Here and in the following sections,
we show, for an LHC energy of
$\sqrt{\spp} = 13\TeV$, 
absolute predictions for $\rd\sigma/\rd\mc{O}$ as
histograms binned in the observable $\mc{O}$  in the upper panel, followed by the relative
corrections of type $i$,
\beq
\label{eq:rel.corrections}
\delta_i(\mc{O}) =
        \frac{\rd\sigma^i}{\rd\mc{O}} \; / \;
        \frac{\rd\sigma^\mr{LO}_{\bar{q}q}}{\rd\mc{O}} ,
\eeq
directly below.

In \reffig{fig:ptes_ATLAS}, we display the transverse-momentum distribution of the electron.
\bfi
\includegraphics[angle=0,scale=0.7,bb=60 330 370 770]{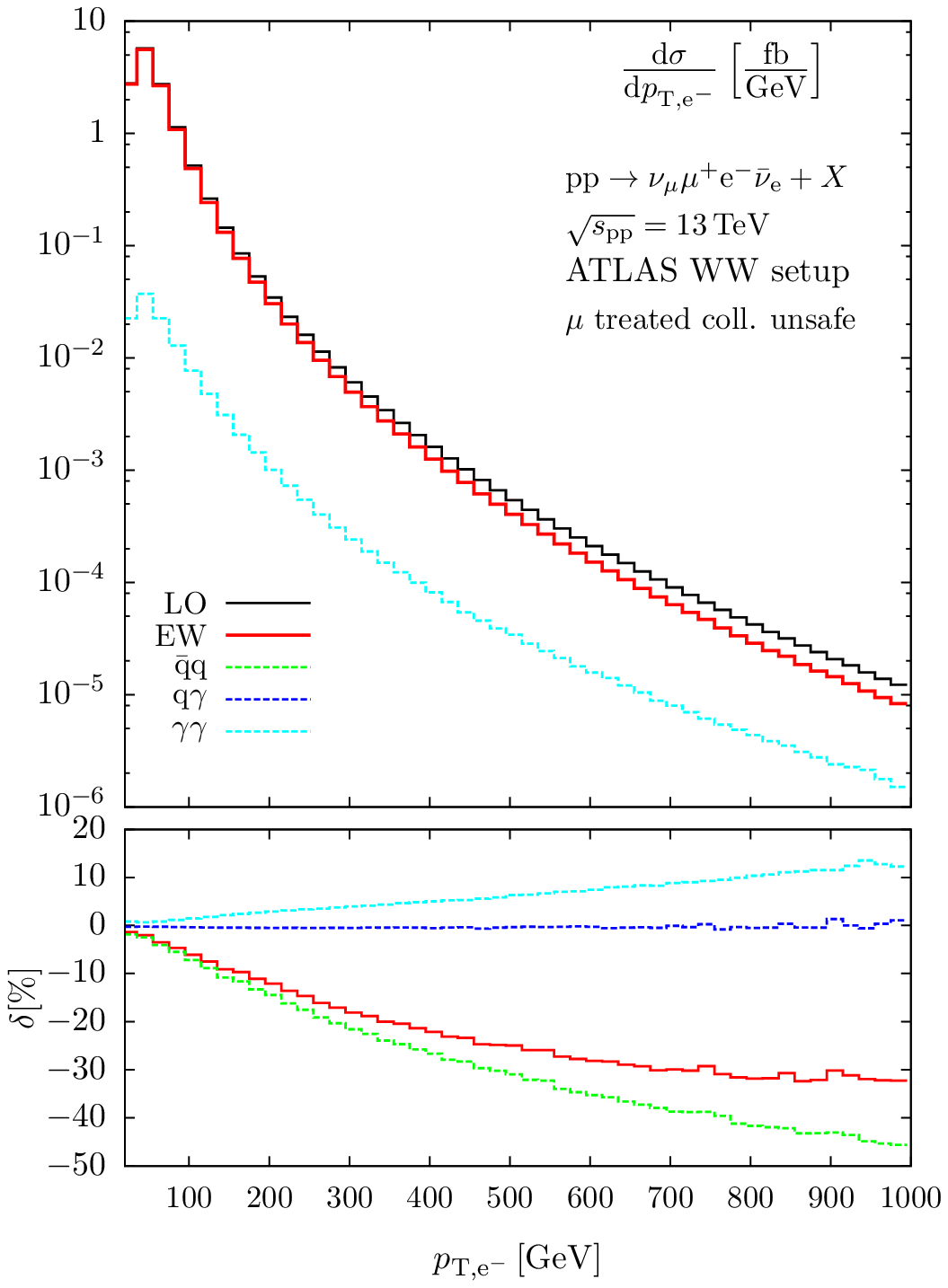}
\includegraphics[angle=0,scale=0.7,bb=60 330 370 770]{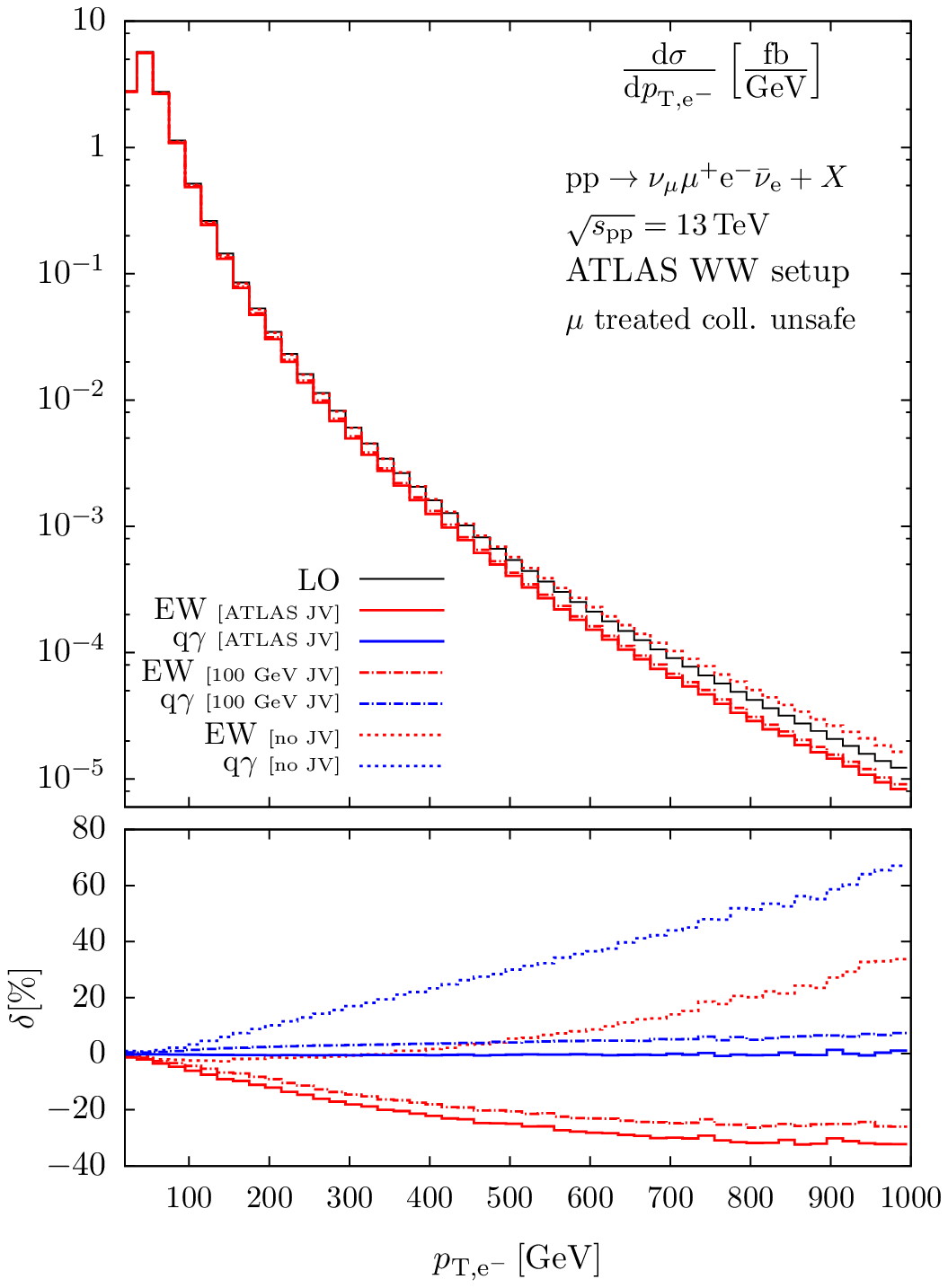}
\caption{Individual contributions to the differential cross section with the default ATLAS jet
veto of $p_{\rT,\mr{jet}}>25\GeV$ (left) and jet-veto (JV) dependence 
(right) of the transverse-momentum distribution of the electron in $\Pp\Pp\to\mvev+X$ in the
ATLAS~WW setup. The lower panels show the relative size of the various corrections.
\label{fig:ptes_ATLAS}
}
\efi
We first concentrate on the left-hand side (l.h.s.)
of the figure.  In the upper
panel, LO refers to $\sigma^\mr{LO}_{\bar{q}q}$ induced by $\bar qq$
channels only, $\gamma\gamma$ to $\sigma^\mr{LO}_{\gamma\gamma}$
induced by photon--photon collisions, and EW to the full NLO EW
prediction, i.e. the sum of the LO cross section with all considered
corrections. 
The $\bar{\Pb}\Pb$-induced contribution to
$\sigma^\mr{LO}_{\bar q q}$ (not shown separately)
is only relevant at low
$p_{\rT,\Pe^-}$, dropping below $2\%$ already at
$p_{\rT,\Pe^-}\sim300\GeV$.  The lower panel compares the relative
corrections induced by the $\bar{q}q$, $q\gamma$, and $\gamma\gamma$
channels as well as their sum (EW).  The corrections in the $\bar{q}q$
channels dominate and exhibit their known negative logarithmic increase
due to weak Sudakov (and subleading) high-energy
logarithms $\propto \alpha/(\pi\sw^2)\ln^2(p_{\rT,\Pe^-}/\MW)$,
reaching $\sim-45\%$ at $p_{\rT,\Pe^-}=1\TeV$.  This huge negative
correction is only partly compensated by the positive contribution of
the $\gamma\gamma$-induced tree-level process, whose impact steadily
grows from about $1\%$ at small transverse momentum to more than $10\%$ at
$1\TeV$.  The contribution of the $q\gamma$-induced channel is
completely insignificant within the ATLAS~WW setup.

The behaviour 
of the $q\gamma$~contribution
is enforced by the jet-identification criterion and the
specific choice of the jet veto within the ATLAS~WW setup,
Eqs.~(\ref{eq:jet-def}) and (\ref{eq:ATLASjet-veto}), respectively.
Both only affect the $q\gamma$-induced contribution, since it 
is the only
contribution that can give rise to a jet.  On the right-hand side (r.h.s.) 
of \reffig{fig:ptes_ATLAS} we illustrate the dependence on the jet veto
(JV) by means of the transverse-momentum distribution
of the electron, where we
show the $q\gamma$-induced contribution together with the full EW
correction for three different values of the jet veto.  If we loosen
the cut of \refeq{eq:ATLASjet-veto} to only reject jets with
$p_{\mathrm{T,jet}} > 100\GeV$ (the value within our inclusive setup,
\refeq{eq:incljet-veto}), the $q\gamma$-induced contribution becomes
positive and leads to a correction of $+0.62\%$
 for the integrated
cross section, resembling the situation in the inclusive setup.
However, the complete omission of a jet veto results in a positive
correction of almost $70\%$ at $p_{\rT,\Pe^-}=1\TeV$, even rendering
the total EW correction $\delta_\mr{EW}$ positive for
$p_{\rT,\Pe^-}\gsim400\GeV$.  In the latter case we obtain a positive
correction of $\sigma_{q\gamma}=3.914(2)\fba$ ($\delta_{q\gamma}=+1.44\%$)
 to the integrated cross section from the
$q\gamma$ channel alone, leading to a total EW correction of only
$\delta_{\EW}=-1.40\%$.
 The reason for this immense increase is a
known mechanism, referred to as ``giant K
factor''~\cite{Rubin:2010xp}, which was already discussed for
$q\gamma$-induced corrections to
W-pair production in the
literature~\cite{Baglio:2013toa,Billoni:2013aba}.  In such processes,
topologies that first occur at NLO introduce kinematic configurations
in which one massive gauge boson may become quasi-soft, leading to
large double-logarithmic corrections for large transverse momenta. In
QCD, such corrections may grow to several $100\%$.  For $q\gamma$
collisions the size is mitigated by the smallness of $\alpha$ and the
photon PDF with respect to $\alpha_{\mathrm{s}}$ and the gluon PDF,
respectively, in spite of some enhancement due to diagrams with
incoming photons coupling to W~bosons, which do not have a QCD
counterpart.
Note also that overwhelmingly large QCD corrections (as a consequence of a missing jet veto)
would force us to calculate EW corrections for
W-pair production in association with hard jets, a task that goes beyond the scope of this paper.

In the following we always apply the jet veto in order to suppress
configurations with hard jets and soft W~bosons, which are less interesting for the investigation
of W-boson pairs.
For the ATLAS~WW and Higgs-background setups, the jet veto actually implies that $q\gamma$-induced 
events may only contribute to the zero-jet cross section, so that any potential final-state jet must 
remain undetected (cf.~\refeq{eq:jet-def}).
This eventually leads to a small negative contribution of the $q\gamma$-induced processes in the 
ATLAS~WW setup of $\sigma_{q\gamma}=-0.744(2) \fba$ ($\delta_{q\gamma}=-0.27\%$), 
the value quoted in \refta{tab:xsecs_extended}.

In Fig.~\ref{fig:mtwwmll_ATLAS} we turn to distributions in the transverse invariant masses of four or two leptons.
\bfi
\includegraphics[angle=0,scale=0.7,bb=60 330 370 770]{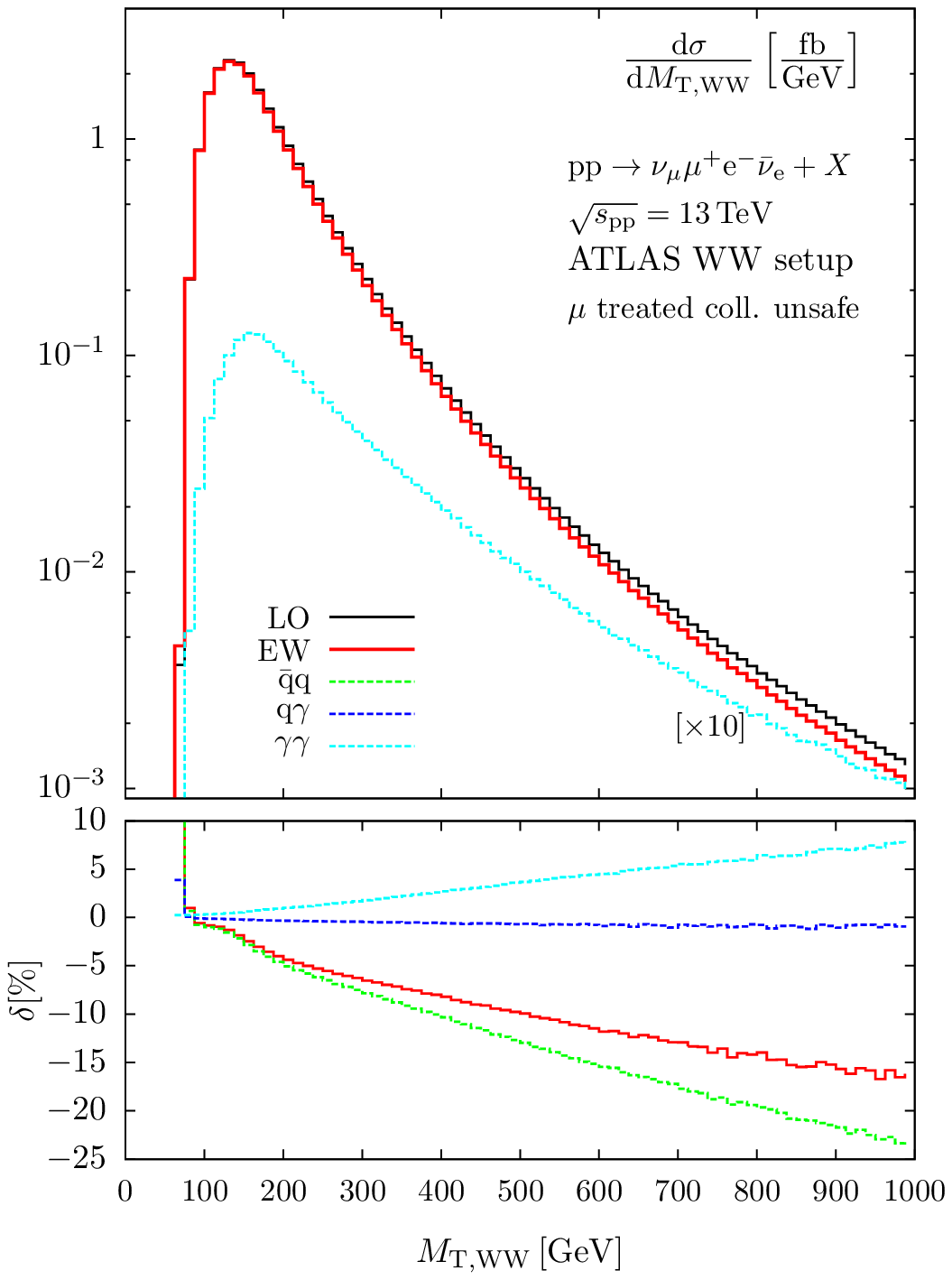}
\includegraphics[angle=0,scale=0.7,bb=60 330 370 770]{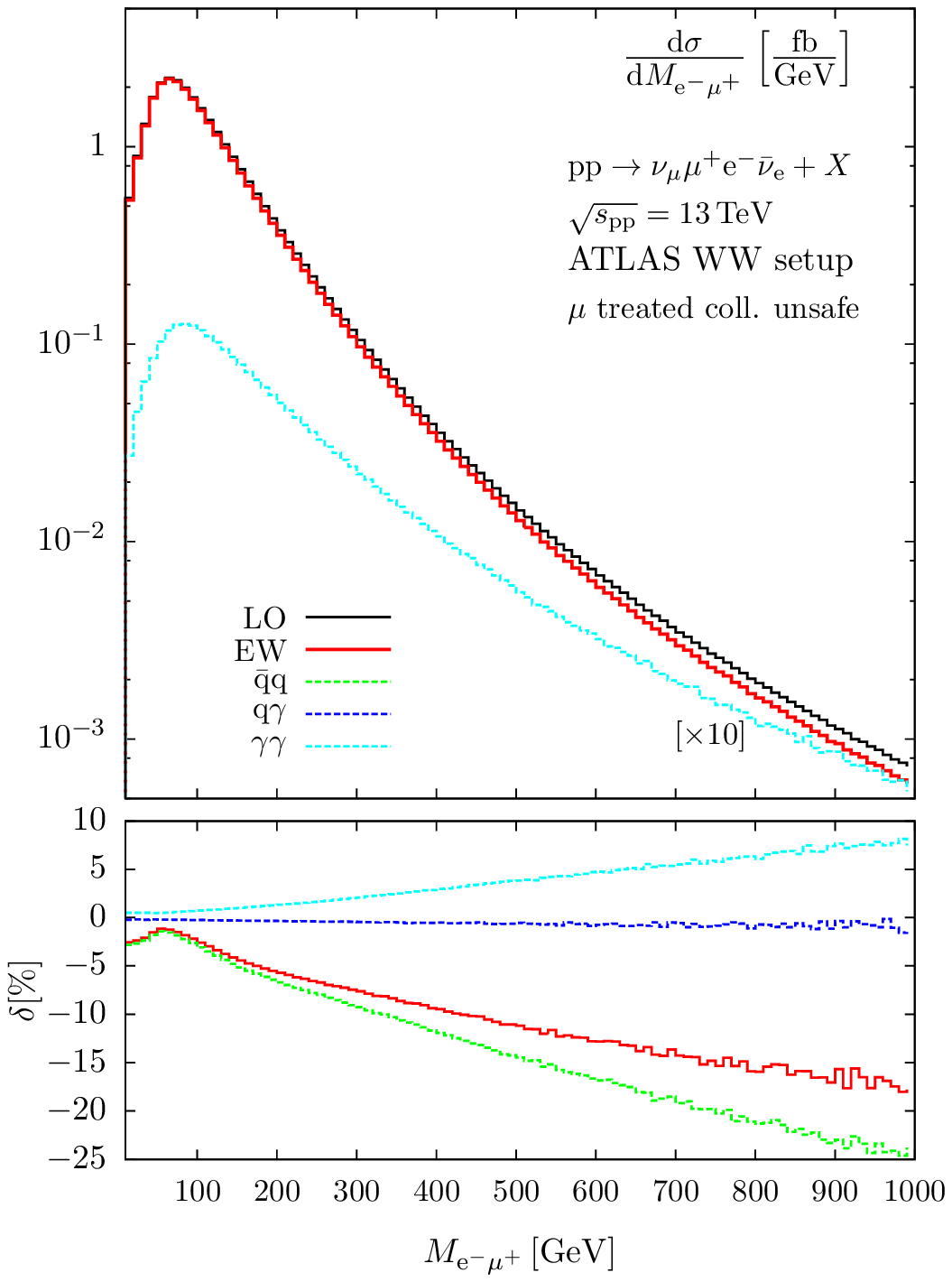}
\caption{Transverse-mass distribution of the four-lepton system (left) and invariant-mass distribution
  of the charged-lepton system (right) in $\Pp\Pp\to\mvev+X$ in the
  ATLAS~WW setup (upper panels), together with the relative impact of
  the individual corrections (lower panels).
  Note that the $\ga\ga$~contribution is scaled by a factor of ten only in the upper panels.
\label{fig:mtwwmll_ATLAS}
}
\efi
Owing to the incomplete information about the momenta of the two neutrinos at a hadron collider,
it is experimentally not possible to fully reconstruct the invariant mass of the 
$\PWp\PWm$ system ($M_{\PW\PW}$) in leptonic final states.
However, in the transverse plane the sum of the neutrino momenta can be inferred from the missing 
transverse momentum ($\vec{p}_\rT^\mr{\;miss}$), motivating the following definition of 
the transverse mass of the four-lepton decay system \cite{Aad:2012uub},
\beq
M_{\rT,\PW\PW} = \sqrt{
(E_{\rT,\Pem\mu^+}+E_{\rT}^{\mr{miss}})^2
-
(\vec{p}_{\rT,\Pem\mu^+}+\vec{p}_{\rT}^{\;\mr{miss}})^2
},
\eeq
with the vector sum 
$\vec{p}_{\rT,\Pem\mu^+}$
of the transverse momenta of the final-state charged leptons, 
the missing transverse momentum 
$\vec{p}_\rT^\mr{\;miss}$ and the corresponding transverse energies given by
\beq
E_{\rT,\Pem\mu^+} = \sqrt{(\vec{p}_{\rT,\Pem\mu^+})^2+M_{\Pem\mu^+}^2} \quad  \textnormal{and}  \quad
E_{\rT}^{\mr{miss}} = |\vec{p}_{\rT}^{\;\mr{miss}}|. 
\eeq
The various contributions to the observable $M_{\rT,\PW\PW}$ are shown in \reffig{fig:mtwwmll_ATLAS} (l.h.s.) 
together with the invariant mass $M_{\Pe^-\mu^+}$ of the charged-lepton system (r.h.s.).
In the high-energy regions, the relative corrections to the two observables exhibit
a very similar quantitative behaviour:
As for the transverse momentum of the electron, the EW correction is dominated by the negative correction 
to the $\bar{q}q$-induced processes.
The positive contribution of the $\gamma\gamma$-induced tree-level process partly compensates for the 
strong negative correction, while the contribution from $q\gamma$ initial states remains insignificant
due to the jet veto.
At the scale of $1\TeV$, we observe a negative total EW correction of about
$-15\%$ for both observables, 
i.e.\ about half the size as in the $p_{\rT,\Pe^-}$ distribution at the same scale.
Note, however, that $\rd\sigma/\rd p_{\rT,\Pe^-}$ 
falls off much steeper with 
$p_{\rT,\Pe^-}$ than the shown invariant-mass distributions with increasing masses. 

In \reffig{fig:etaedeltaphill_ATLAS} we show rapidity and angular distributions of the charged leptons 
within the ATLAS~WW setup.
\bfi
\includegraphics[angle=0,scale=0.7,bb=60 330 370 770]{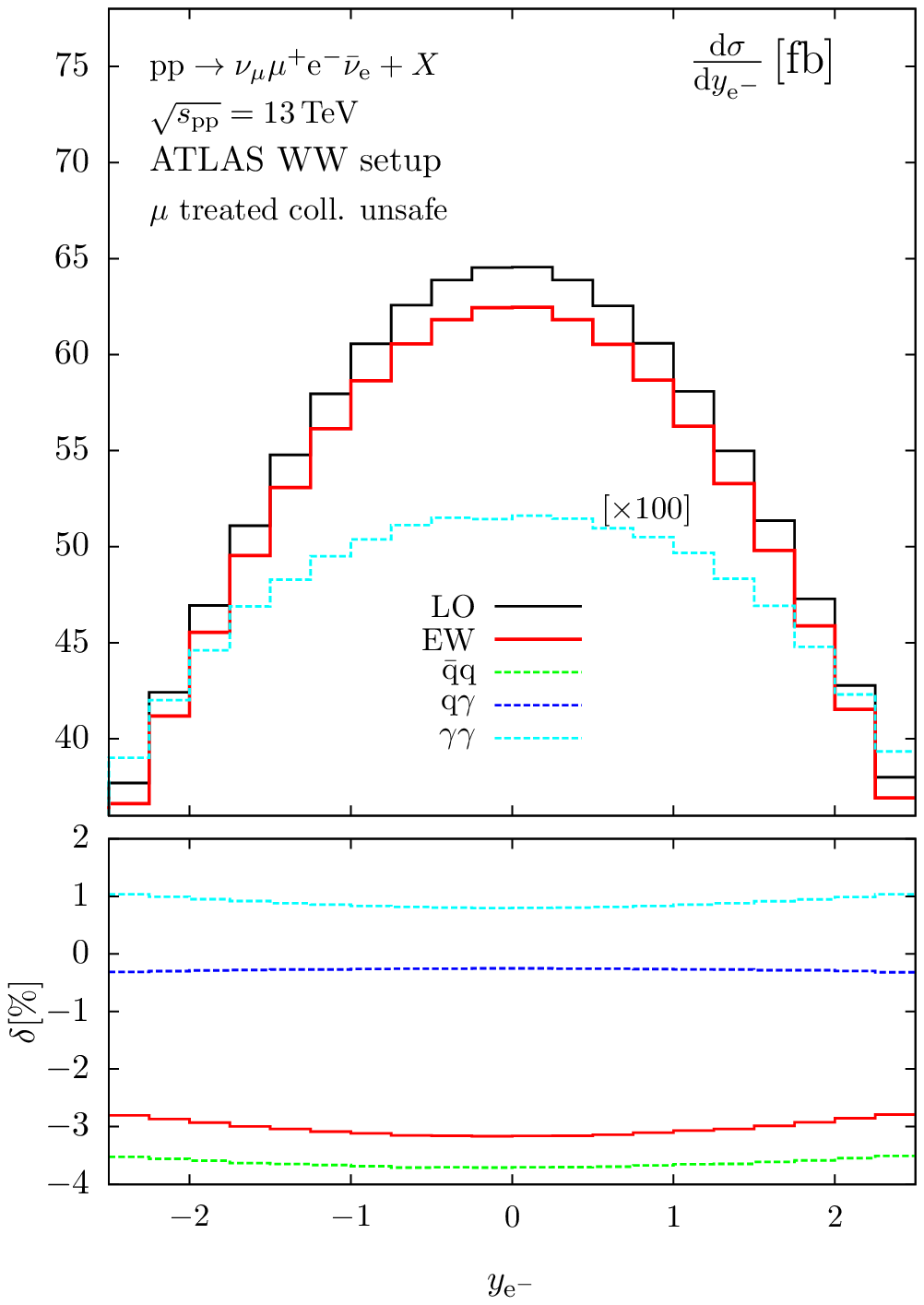}
\includegraphics[angle=0,scale=0.7,bb=60 330 370 770]{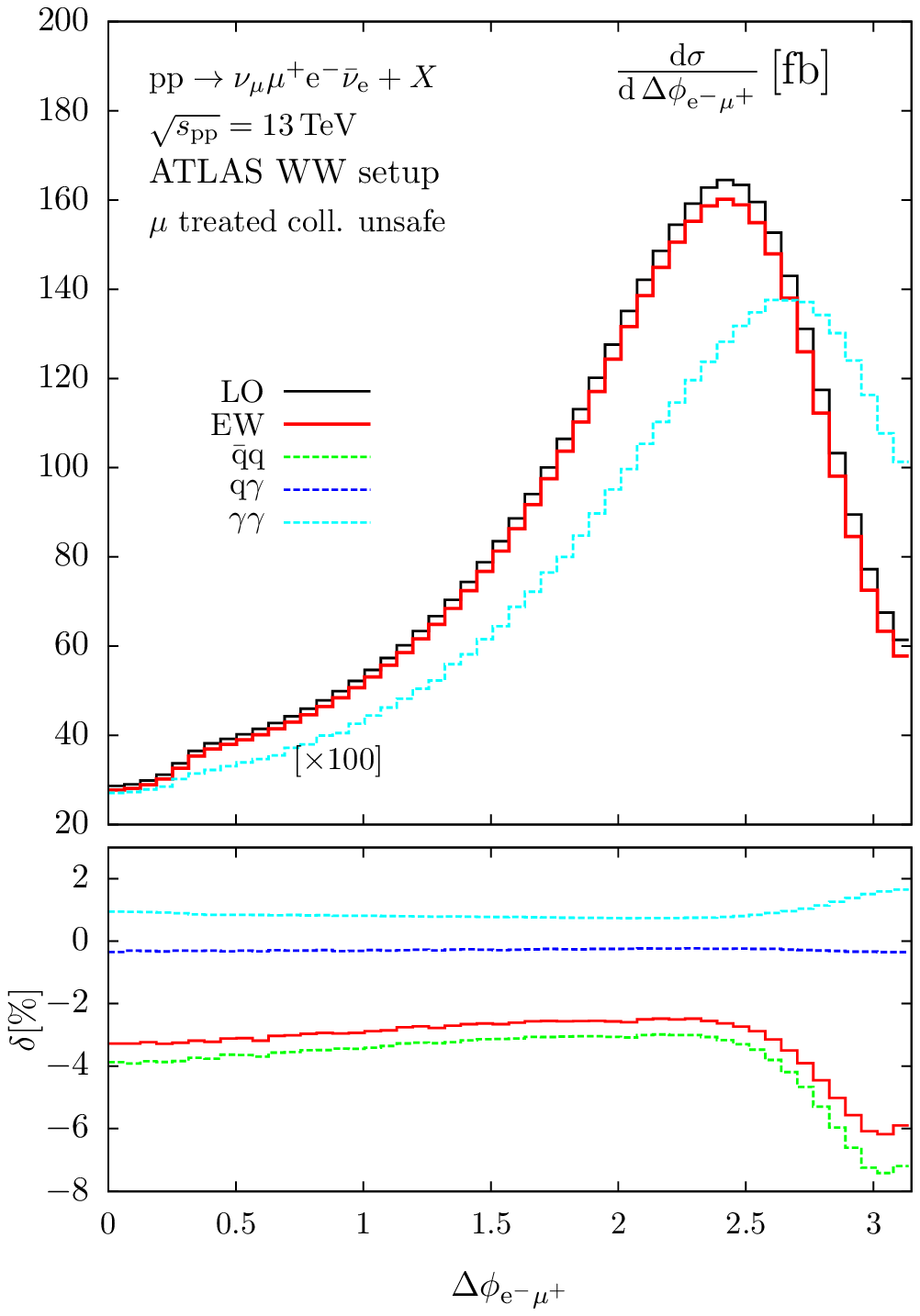}
\caption{Rapidity distribution of the electron (left) and distribution in the
azimuthal-angle separation of the two charged leptons 
(right) in $\Pp\Pp\to\mvev+X$ in the ATLAS~WW setup.
The lower panels show the relative impact of the various contributions.
Note that the $\ga\ga$~contribution is scaled by a factor of hundred only in the upper panels.
\label{fig:etaedeltaphill_ATLAS}
}
\efi
The corrections to the rapidity distribution of the electron (left)
are uniformly distributed and resemble the corrections to the
integrated cross sections given in \refta{tab:xsecs_extended}.  The
azimuthal-angle separation of the two charged leptons (right), on the
other hand, receive some distortion due to the EW corrections towards
a separation of $\Delta\phi_{\Pem\mu^+}=\pi$ of the two charged
leptons.  This back-to-back configuration in the transverse plane is
favoured by events with $\PW$-boson pairs with large transverse
momenta, which explains the tendency to receive more negative EW
corrections in the $\bar qq$ channels.

\FloatBarrier

\subsection{Results on differential cross sections in the Higgs-background setup}

A very important irreducible background to the decay of a Higgs boson
to a $\PW$-boson pair, $\PH \to \PW\PW^*$, originates from the direct
\PW-pair production process $\Pp\Pp\to \PW\PW^*\to \mvev +X$.  In the
following, we study the influence of the EW corrections on this
dominant 
background in the Higgs signal region defined by the
additional cuts of \refeq{eq:HtoWWcut} and \refeq{eq:HtoWWcut2}.
Since this setup is meant to favour the Higgs signal and to suppress
any background as much as possible, we observe a reduction of
$\sigma^\mr{LO}_{\bar q q}$ by almost a factor of six compared to the
ATLAS~WW setup (cf.~\refta{tab:xsecs_extended}).  We also observe some
reduction of all relative corrections, resulting in a total EW
correction of only $-2.25\%$ for the $13\TeV$ prediction (we again
provide results for $13\TeV$ and the collinear-unsafe photon
scenario).  This reduction can be explained by
looking at the differential distributions of the two observables to
which the cuts of \refeq{eq:HtoWWcut} are applied.

In \reffig{fig:mlldeltaphill_HtoWW} we show these observables, namely
the invariant mass (left) and the azimuthal-angle separation (right)
of the two charged leptons, within the Higgs-background setup.
\bfi
\includegraphics[angle=0,scale=0.7,bb=60 330 370 770]{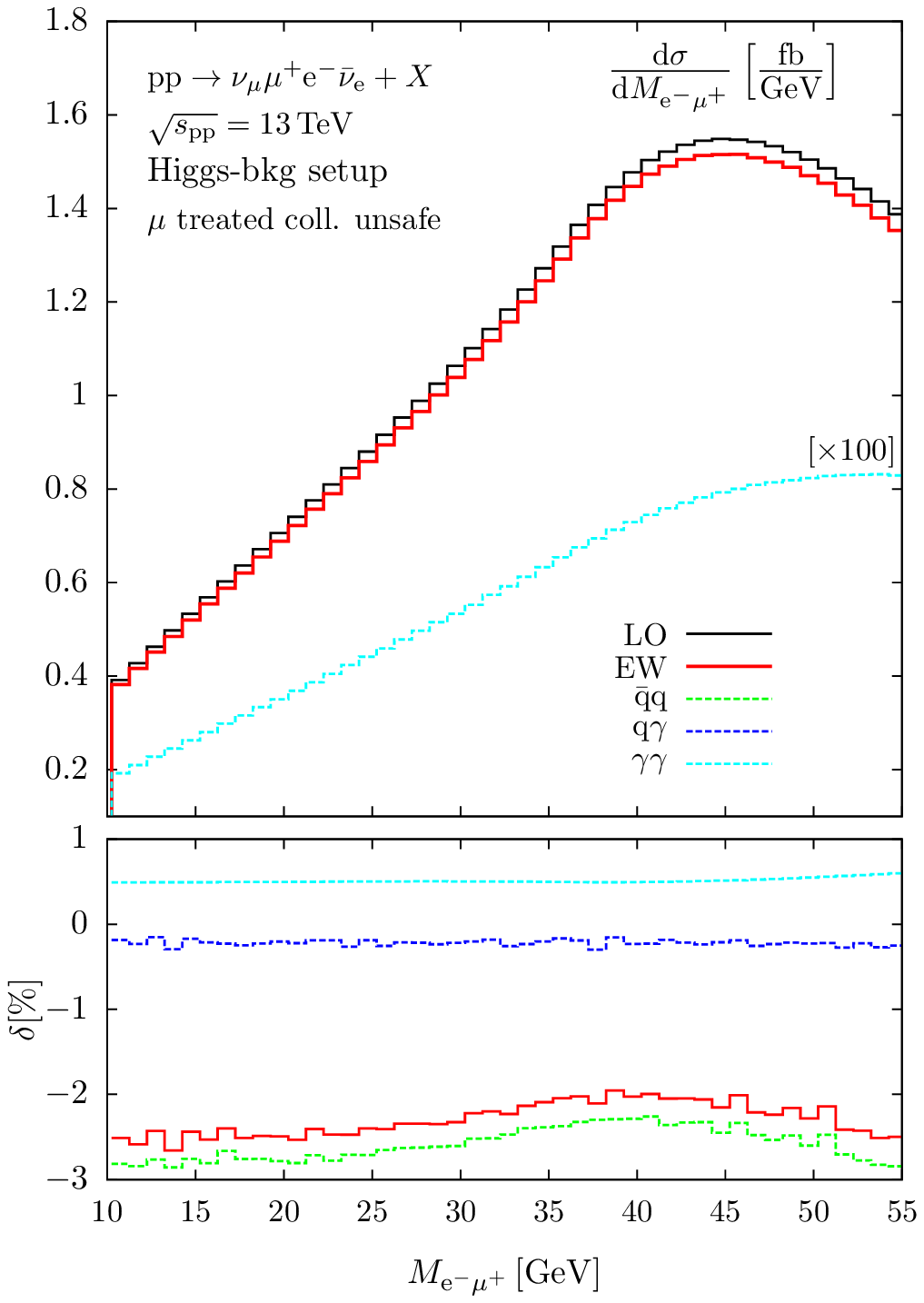}
\includegraphics[angle=0,scale=0.7,bb=60 330 370 770]{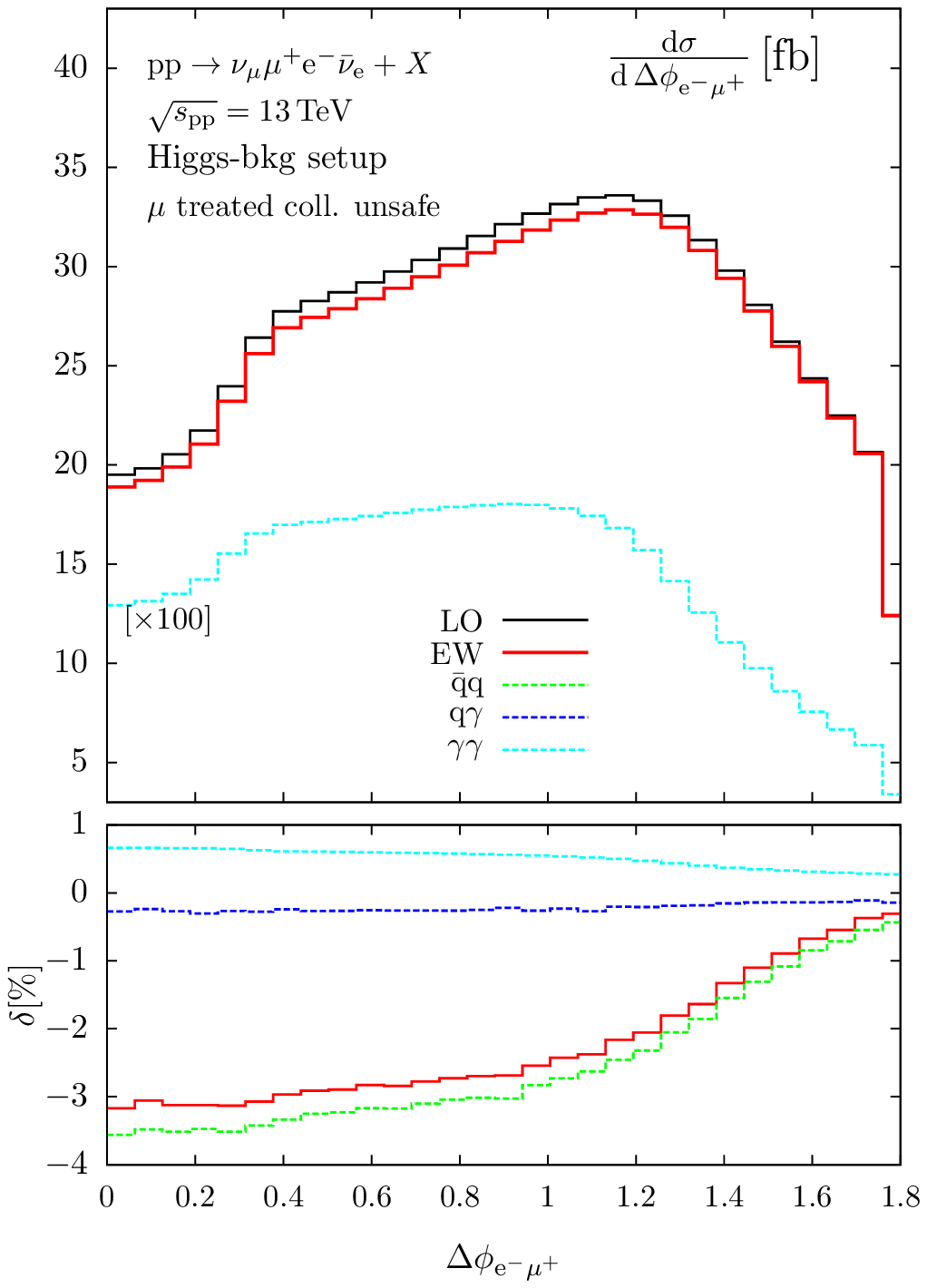}
\caption{Distributions in the invariant mass of the charged-lepton system (left) and in the 
azimuthal-angle separation of the two charged leptons (right) in $\Pp\Pp\to\mvev+X$ in the Higgs-background setup.
The lower panels show the relative impact of the various contributions.
Note that the $\ga\ga$~contribution is scaled by a factor of hundred only in the upper panels.
\label{fig:mlldeltaphill_HtoWW}
}
\efi
As discussed above, the large relative corrections to the invariant-mass distribution of the two charged leptons
in the ATLAS~WW setup were observed for large $M_{\Pe^-\mu^+}$, a region that is completely removed by the 
additional cuts, so that smaller EW corrections are expected.
In the allowed range of the invariant mass of the charged-lepton system we now observe quite uniformly 
distributed corrections from all contributions (\reffig{fig:mlldeltaphill_HtoWW}, left).
For the azimuthal-angle separation (\reffig{fig:mlldeltaphill_HtoWW}, right),
the region of phase space exhibiting the most pronounced EW corrections in the ATLAS~WW setup has been cut away, 
but the cuts affect the LO distribution and the corrections in the allowed range in a non-trivial way. 
Towards the new maximal value of $\Delta\phi_{\Pe^-\mu^+}^\mr{cut}=1.8$ we observe a strong decrease of the 
cross section and a reduction of the EW corrections.

The transverse-momentum distribution of the electron and the transverse-mass distribution of the \PW-pair 
in the Higgs-background setup are shown in \reffig{fig:ptemtww_HtoWW}. 
\bfi
\includegraphics[angle=0,scale=0.7,bb=60 330 370 770]{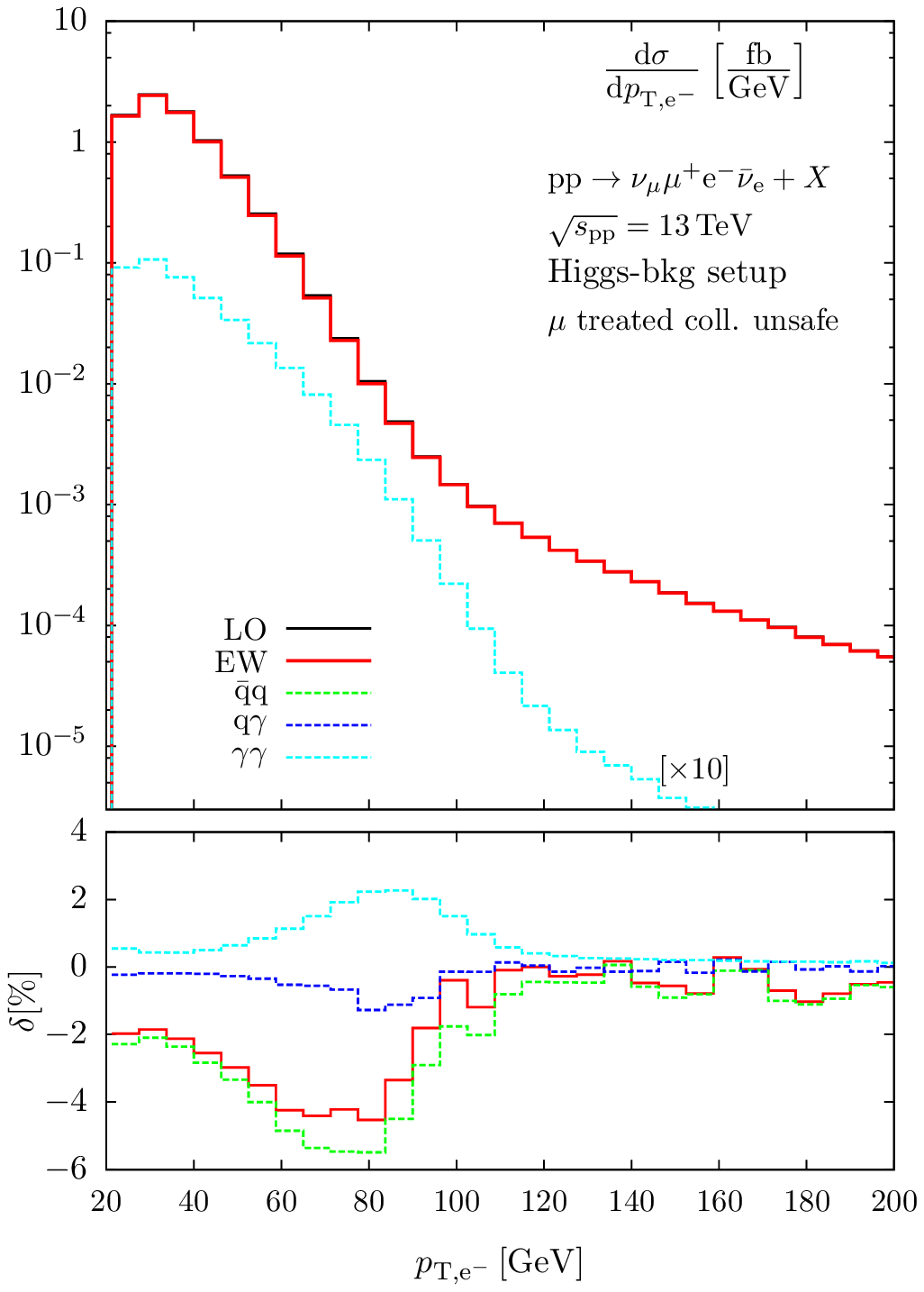}
\includegraphics[angle=0,scale=0.7,bb=60 330 370 770]{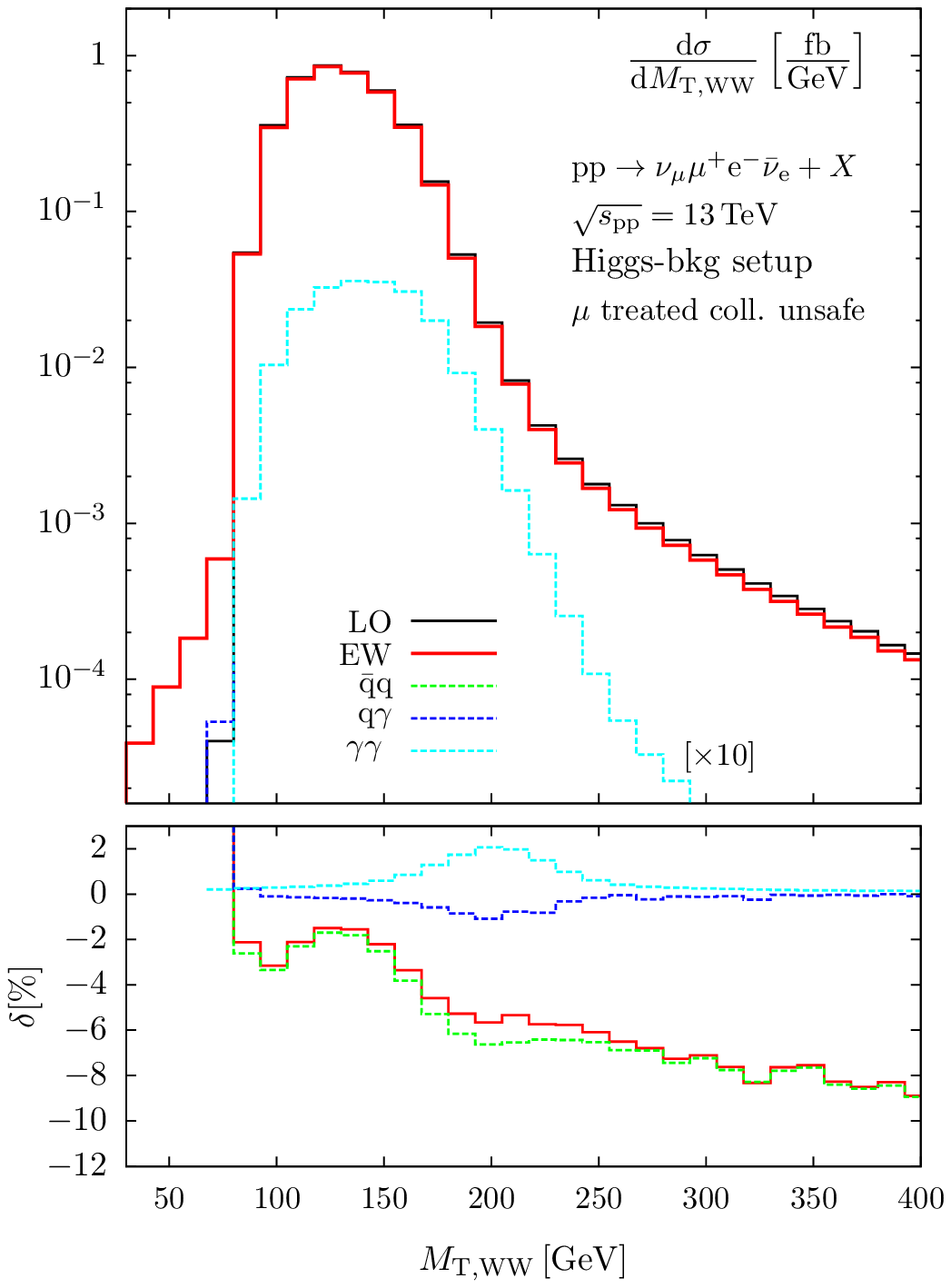}
\caption{Transverse-momentum distribution of the electron (left) and transverse-mass distribution of the four-lepton system (right) in $\Pp\Pp\to\mvev+X$ in the
Higgs-background setup. The lower panels show the relative impact of the various contributions.
Note that the $\ga\ga$~contribution is scaled by a factor of ten only in the upper panels.
\label{fig:ptemtww_HtoWW}
}
\efi
We point out that both observables exhibit a much steeper decrease of
the LO cross section in the shown kinematic range than within the
ATLAS~WW setup (cf.~\reffigs{fig:ptes_ATLAS} and
\ref{fig:mtwwmll_ATLAS}): The distributions in
the Higgs setup drop faster with increasing scales by roughly a factor
of $100$ compared to the situation in the ATLAS~WW setup.  The
corrections induced by the $q\gamma$ and $\gamma\gamma$ channels
almost cancel each other at low
and are suppressed at large
scales.  The EW corrections are 
thus almost entirely due to
corrections to the $\bar qq$ channels.  They distort the shapes of the
distributions significantly in a non-trivial way.  While the EW
corrections to the $M_{\rT,\PW\PW}$ distribution
(\reffig{fig:ptemtww_HtoWW}, right) show the onset of the typical
decrease towards larger scales, the EW corrections to the
$p_{\rT,\Pe^-}$ distribution are significant only for
$p_{\rT,\Pe^-}\lsim100\GeV$.

In \reffig{fig:mwws_HtoWW} we investigate the (experimentally unobservable) invariant-mass distribution of the 
four-lepton system, where the Higgs-boson resonance is located at $M_{\PW\PW}=\MH\approx125\GeV$.
\bfi
\includegraphics[angle=0,scale=0.7,bb=60 330 370 770]{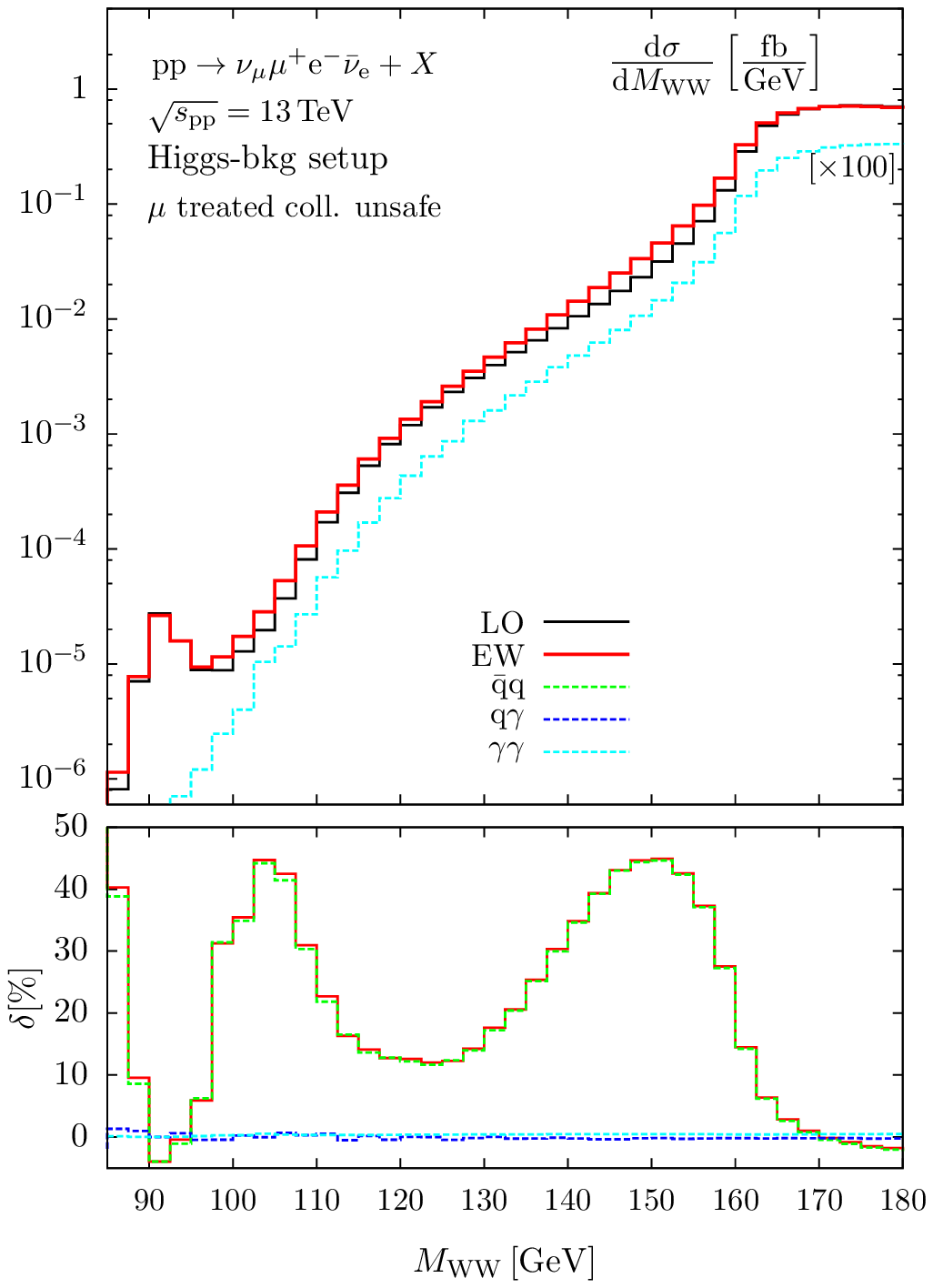}
\includegraphics[angle=0,scale=0.7,bb=60 330 370 770]{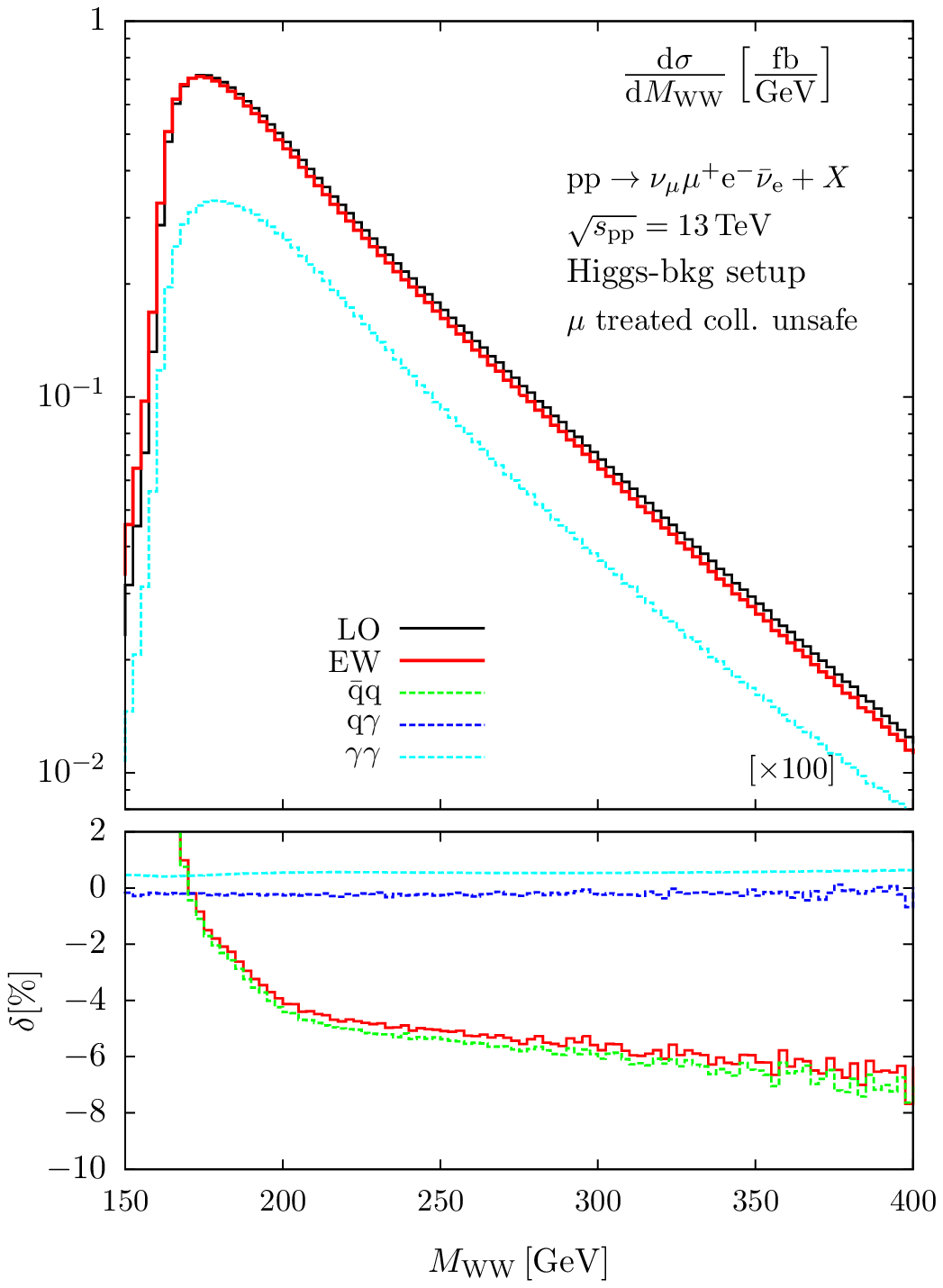}
\caption{Invariant-mass distribution of the four-lepton system in two different plot ranges in 
$\Pp\Pp\to\mvev+X$ in the Higgs-background setup. 
The lower panels show the relative impact of the various contributions.
Note that the $\ga\ga$~contribution is scaled by a factor of hundred only in the upper panels.
\label{fig:mwws_HtoWW}
}
\efi
Between $M_{\PW\PW}=80\GeV$ and the on-shell \PW-pair threshold at
$M_{\PW\PW}=2\MW\approx160\GeV$, we observe a very strong increase by
almost five orders of magnitude in the $M_{\PW\PW}$~distribution.
Although we clearly see that the direct production of a \PW-boson pair
within the Higgs-background setup is still dominated by on-shell
W-pairs with $M_{\PW\PW}\gsim2\MW$, it is still interesting to look
into the region below this threshold, where at least one of the
W~bosons is forced to be off shell.  At $M_{\PW\PW}=M_\PZ$, 
the \PZ-boson resonance is visible, 
though very strongly suppressed, since at least one of the \PW~bosons
has to be far off shell
 there.  The distinct structures and the strong
positive enhancement of the EW corrections below the W-pair threshold
can be attributed to the kinematic redistribution of events by
collinear final-state radiation of photons off the charged leptons.
This effect systematically shifts events to lower values of
$M_{\PW\PW}$, leading to pronounced positive corrections where the
spectrum falls off steeply with decreasing values of $M_{\PW\PW}$.
This well-known feature near kinematical thresholds has recently also
been discussed in a similar setup for the EW corrections for off-shell
\PZ-pair production~\cite{Biedermann:2016yvs}.  At the invariant
\PW-pair mass of the order of the Higgs-boson mass of $125\GeV$ we
observe a positive EW correction of about $+15\%$, but we remind the
reader that $M_{\PW\PW}$ is not an observable for purely leptonically
decaying W-boson pairs, in contrast to the respective situation for
Z-boson pairs.  Above the WW threshold, where resonant W~bosons
dominate, the EW correction shows again the tendency to grow into the
negative direction with increasing $M_{\PW\PW}$.  We point out,
however, that this growth is much slower than observed in the
transverse-momentum and transverse-mass distributions
(in particular in the ATLAS WW setup, cf.~Figs.~\ref{fig:ptes_ATLAS}
and \ref{fig:mtwwmll_ATLAS}), because the
region of large invariant mass $M_{\PW\PW}$ 
is dominated by forward-scattered W~bosons owing to $t$-channel diagrams,
and thus not by the
Sudakov regime where all momentum invariants have to be large.
Finally, we note that corrections due to the photon-induced channels
do not play a significant role in this distribution.

\FloatBarrier

\subsection{Collinear-safe versus collinear-unsafe case}
\label{ssec:safevsunsafe}

In this section we discuss the impact of the recombination of nearly collinear photons with final-state 
leptons on total and differential cross sections.
In all results shown up to now, the recombination procedure described at the end of \refse{ssec:input}
was applied only to electrons, while muons were treated in a collinear-unsafe way. 
In \refta{tab:xsecs_safeunsafe} we list $\sigma^\mr{LO}_{\bar q q}$ for our three phenomenological cut 
scenarios at the LHC operating at an energy of $\sqrt{\spp}=13\TeV$, together with the corrections 
from the $\bar{q}q$-induced channels for our default setup (collinear-unsafe case) and for the case 
where we apply the recombination procedure to both charged leptons (collinear-safe case).
\begin{table}
\begin{center}
\begin{tabular}{|c|c|cc|}
\hline
{\small{LHC $\; 13\TeV$}} & $\sigma^\mr{LO}_{\bar q q}$~[fb] &  $\delta_{\bar q q}^{\mr{coll.~unsafe}}$~[$\%$]&  $\delta_{\bar q q}^{\mr{coll.~safe}}$~[$\%$]
\\
\hline
\hline

inclusive  &390.59(3)& $-3.41$   & $-2.91$  \\
\hline
\hline
ATLAS~WW  &271.63(1) & $-3.71$ &  $-3.18$  \\
\hline
\hline
Higgs~bkg  &$49.934(2)$  & $-2.54$  & $ -1.95$ \\
\hline
\end{tabular}
\caption{
\label{tab:xsecs_safeunsafe}
LO cross sections for $\Pp\Pp\to \mvev+X$ at the LHC running at $13\TeV$
in the inclusive cut scenario (top line), the ATLAS~WW setup (middle line), 
and the Higgs-background setup (bottom line). In the last two columns we list the relative corrections to the $\bar q q$-induced contributions in our default setup (collinear unsafe) and the collinear-safe setup. 
The numbers in brackets represent the numerical error on the last given digit.  
}
\end{center}
\end{table}
For integrated cross sections we observe slightly reduced corrections
(by about $0.5\%$) in all three cut scenarios 
in the fully collinear-safe setup.  This is due to the missing enhancement of
final-state radiation by the mass singularity that appears in the
collinear-unsafe treatment of muons.  
In the fully collinear-safe case
more muons pass the cuts (after recombination with photons),
so that the correction tends to be less negative.

This effect can most directly be
observed in the transverse-momentum distribution of the muon
and in the W-pair invariant-mass distribution, which are both shown in \reffig{fig:ptmumww_diffs_ATLAS}
in the ATLAS~WW setup.
\bfi
\includegraphics[angle=0,scale=0.7,bb=60 330 370 770]{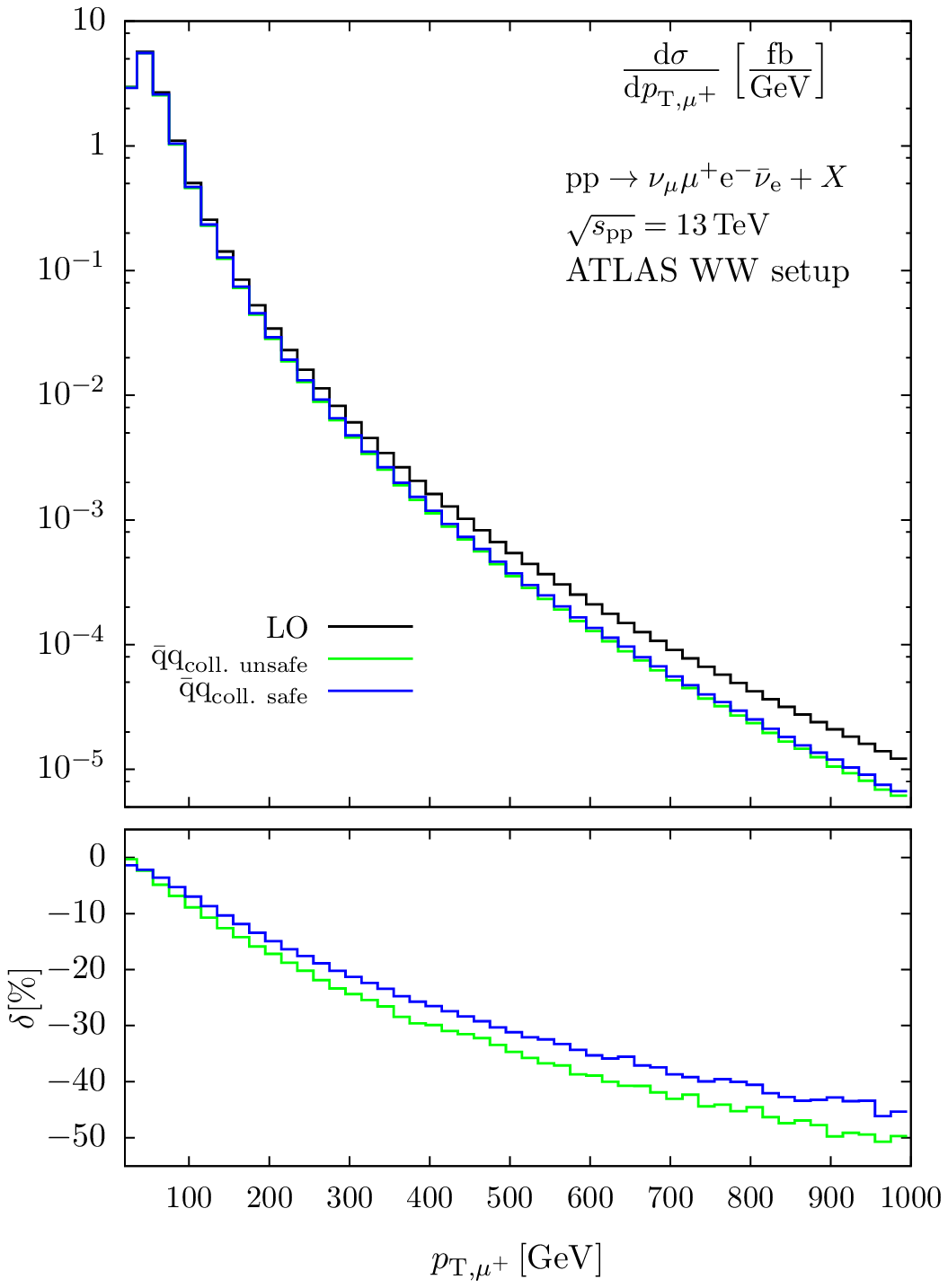}
\includegraphics[angle=0,scale=0.7,bb=60 330 370 770]{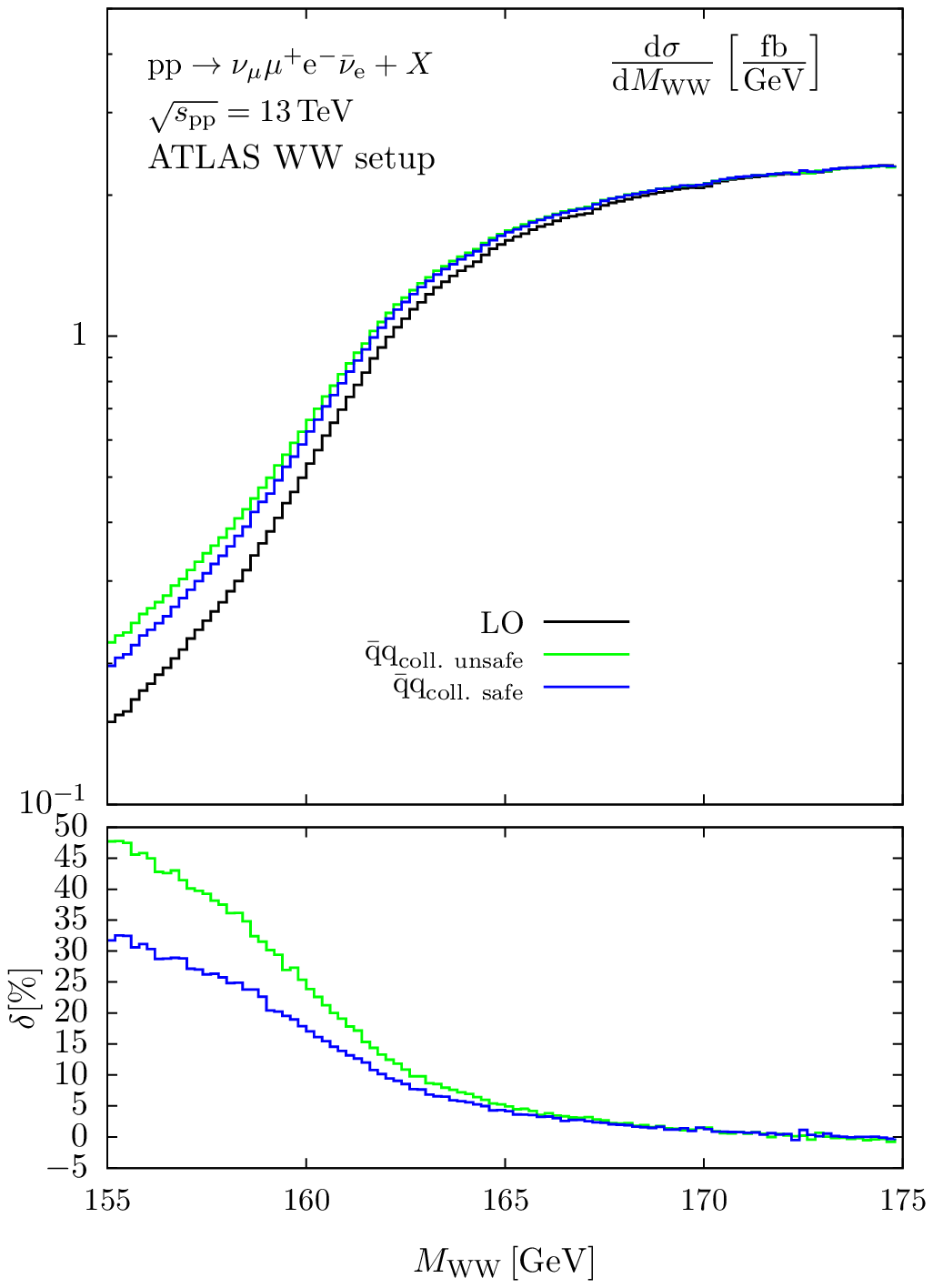}
\caption{Transverse-momentum distribution of the muon (left) and invariant-mass distribution of the 
four-lepton system (right) in $\Pp\Pp\to\mvev+X$ in the
ATLAS~WW setup. The lower panels show the relative size of the $\bar{q}q$ contribution within our 
default (collinear-unsafe) setup compared to the collinear-safe case.
\label{fig:ptmumww_diffs_ATLAS}
}
\efi
In the $p_{\rT,\mu^+}$ distribution, which is widely monotonically falling, the collinear-unsafe 
correction factor signals larger negative corrections than its collinear-safe counterpart over the
entire plot range, apart from the first bins where the maximum of the distribution is located.
In the $M_{\PW\PW}$ distribution the converse situation is observed:
For $M_{\PW\PW}\gsim165{-}175\GeV$, where the distribution is rather flat, hardly any difference between
collinear-unsafe and collinear-safe event selection is visible, because events are shifted more or less uniformly
by photon recombination. 
For $M_{\PW\PW}\lsim2\MW$, however, much more events migrate from larger to smaller invariant masses
in the collinear-unsafe case because of the mass-singular enhancement of final-state radiation, leading
to larger positive corrections as compared to the collinear-safe case.

Angular distributions in the final-state leptons (not shown)
do not exhibit significant distortions when applying 
the recombination procedure, since recombining leptons and collinear photons does not systematically
change the lepton direction.
The difference between the two treatments directly reflects the difference observed in the integrated 
results over the full phase space. 

\FloatBarrier

\subsection{\bf Comparison to the double-pole approximation} 
\label{ssec:fullvsdpa}

In this section we discuss the validity and quality of the DPA,
which was constructed for the virtual EW corrections to hadronic W-pair production in \citere{Billoni:2013aba},
by comparing integrated and differential results in DPA to results of our full $4f$~calculation. 
In \refta{tab:xsecs_DPA} we list the LO cross sections $\sigma^\mr{LO}_{\bar q q}$ and 
$\sigma^\mr{LO,DPA}_{\bar q q}$, for our three setups at the LHC at an energy of $\sqrt{\spp}=13\TeV$. 
\begin{table}
\begin{center}
\begin{tabular}{|c|cc|cc|}
\hline
{\small{LHC $\; 13\TeV$}} &$\sigma^\mr{LO}_{\bar q q}$~[fb]&$\sigma^\mr{LO, DPA}_{\bar q q}$~[fb] &  $\delta_{\bar q q}$~[$\%$]&  $\delta_{\bar q q}^{\mr{DPA}}$~[$\%$]
\\
\hline
\hline
inclusive  &390.59(3)&384.96(9)   & $-3.41$   & $-3.43$  \\
\hline
\hline
ATLAS~WW  &271.63(1)& 265.31(3)   & $-3.71$  & $ -3.68$  \\
\hline
\hline
Higgs~bkg  &49.934(2)&  48.88(2)  & $-2.54$   & $ -2.54$   \\
\hline
\end{tabular}
\caption{
\label{tab:xsecs_DPA}
LO cross sections for $\Pp\Pp\to \mvev+X$ at the LHC running at $13\TeV$ 
in the inclusive setup (top line), the ATLAS~WW setup (middle line), and the Higgs-background setup (bottom line). 
In the last two columns we list the relative EW corrections to the $\bar q q$-induced contributions including the full virtual corrections ($\delta_{\bar q q}$) and applying the DPA within the virtual contributions ($\delta^\mr{DPA}_{\bar q q}$), both normalized to $\sigma^\mr{LO}_{\bar q q}$. 
The numbers in brackets represent the numerical error on the last given digit.
}
\end{center}
\end{table}
The cross section 
$\sigma^\mr{LO,DPA}_{\bar q q}$
includes only doubly-resonant diagrams for the $\bar{q}q$-induced contributions.
The difference of approximately $2\%$ indicates that non-doubly-resonant contributions only contribute 
at the expected level of ${\cal O}(\Gamma_\PW/\MW)$, which is, however, not good enough to achieve
percent-level accuracy even after including higher-order corrections.
Note that we include the DPA LO cross section $\sigma^\mr{LO,DPA}_{\bar q q}$ in our discussion for illustration
only, but that the full $4f$~LO cross section $\sigma^\mr{LO}_{\bar q q}$ was consistently used
in the evaluation of the DPA NLO cross section in \citere{Billoni:2013aba}.
As discussed already in \refse{sec:4fvsdpa}, we define the relative corrections of the full 
$4f$~prediction and the DPA upon normalizing to the full LO cross section,
\beq
 \delta_{\bar{q}q}(\mc{O}) =
        \frac{\rd\sigma^\mr{NLO}_{\bar{q}q}}{\rd\mc{O}} \;/\;
        \frac{\rd\sigma^\mr{LO}_{\bar{q}q}}{\rd\mc{O}} \,,
\qquad 
 \delta_{\bar{q}q}^{\mr{DPA}}(\mc{O}) =
        \frac{\rd\sigma_{\bar{q}q}^{\mr{NLO,DPA}}}{\rd\mc{O}} \;/\;      
        \frac{\rd\sigma^\mr{LO}_{\bar{q}q}}{\rd\mc{O}} \,,
\eeq   
so that they only differ within the virtual contributions of the NLO calculation.
For integrated cross sections, the two corrections show very good agreement, 
as can be seen in the rightmost columns 
of \refta{tab:xsecs_DPA}.

A similar observation can be made for the rapidity distributions of the leptons,
which is illustrated for the one of the electron on the l.h.s.\ of \reffig{fig:etaemll_DPA_ATLAS}
in the ATLAS~WW setup.
\bfi
\includegraphics[angle=0,scale=0.7,bb=60 330 370 770]{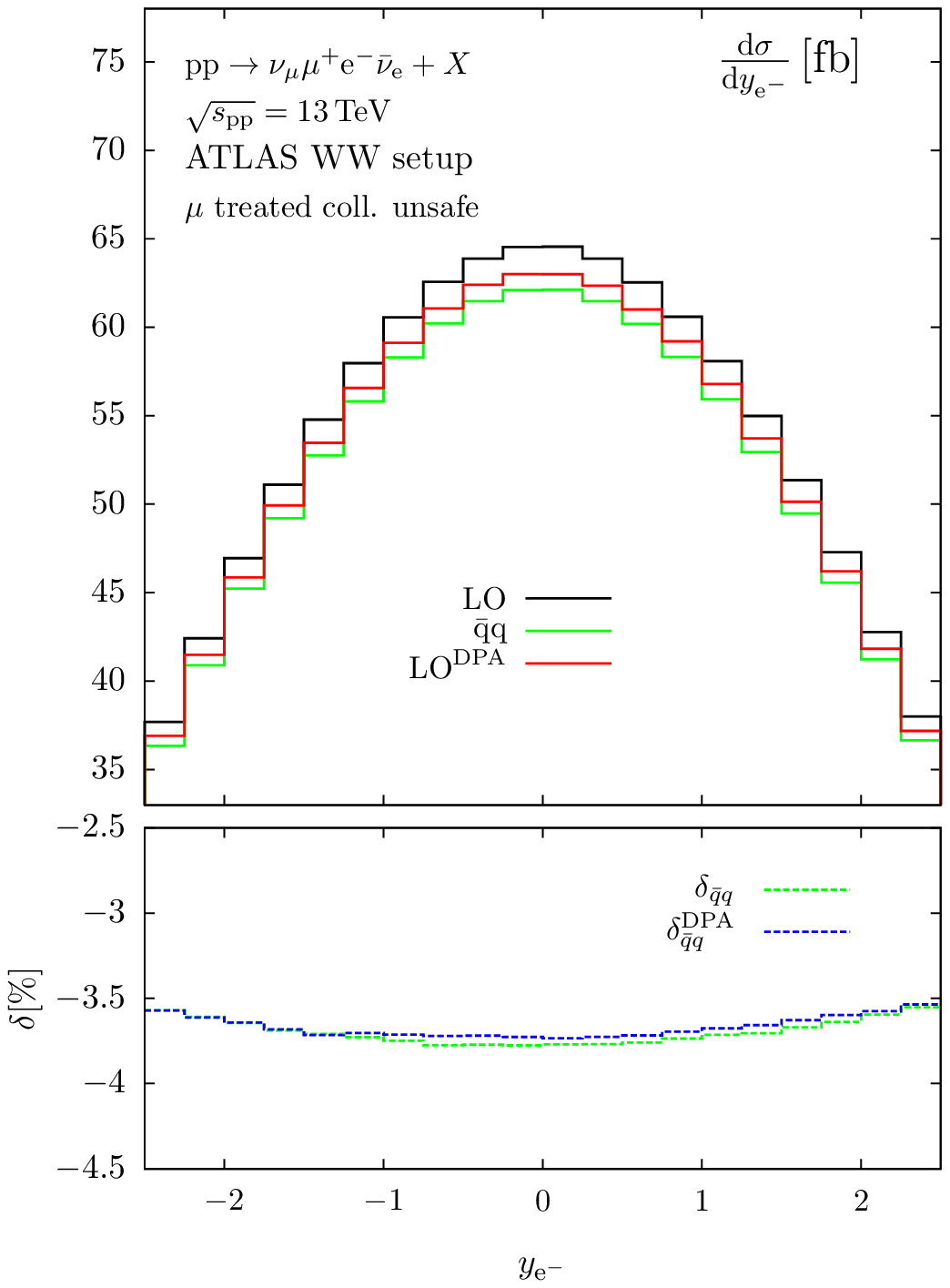}
\includegraphics[angle=0,scale=0.7,bb=60 330 370 770]{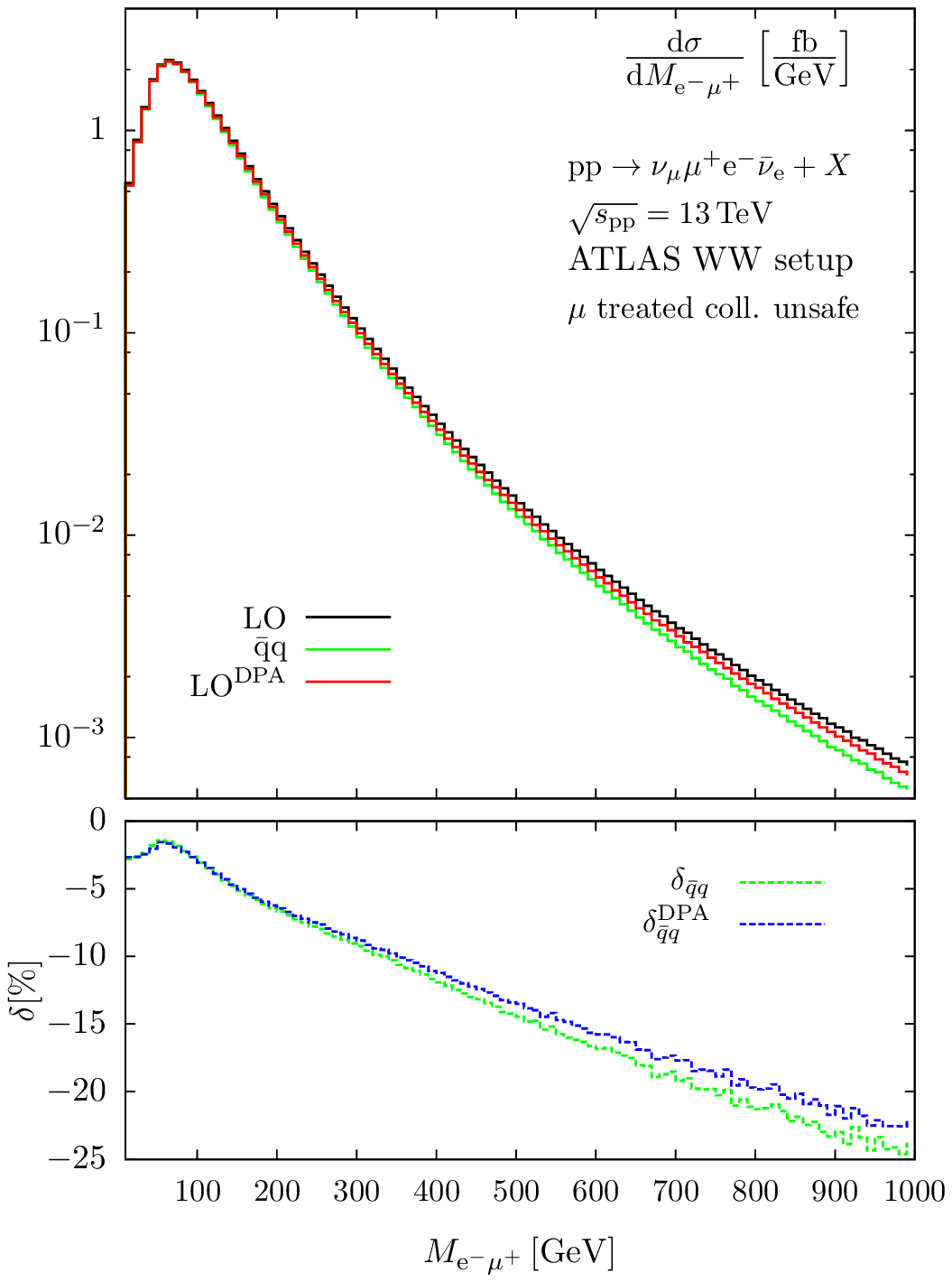}
\caption{Rapidity distribution of the  electron (left) and invariant-mass distribution of the charged-lepton system (right) in $\Pp\Pp\to\mvev+X$ in the
ATLAS~WW setup. The lower panels show the relative size of the EW corrections to the $\bar{q}q$ channels
in our default setup compared to the result based on the DPA.
\label{fig:etaemll_DPA_ATLAS}
}
\efi
In the upper panel we observe the clear deviation of $\sigma^\mr{LO,DPA}_{\bar q q}$ from the full LO 
prediction, being of the same order of magnitude as the EW corrections to the $\bar q q$-induced processes.
The lower panel shows the excellent agreement of the two versions for the relative corrections, with
differences at the $0.1\%$ level only.

The r.h.s.\ of \reffig{fig:etaemll_DPA_ATLAS} illustrates the same
comparison for the invariant-mass distribution of the charged-lepton
system in the ATLAS~WW setup.  For $M_{\Pem\mu^+}\lsim500\GeV$, the
DPA is accurate 
within $1\%$, but the difference grows to about $2{-}3\%$
in the TeV range. This increasing difference between the full
$4f$~calculation and the DPA can already be inferred
from the LO cross sections $\sigma^\mr{LO}_{\bar q
  q}$ and $\sigma^\mr{LO,DPA}_{\bar q q}$ in the upper panel, which
signals the increasing impact of singly-resonant contributions 
not being included in the DPA.  The difference between full and DPA NLO
EW corrections is well covered by the last term of our estimate
\refeq{eq:DeltaDPA}.  In view of the typically expected accuracies in
LHC data analyses, the DPA is certainly sufficient for this
observable.

Finally, in \reffig{fig:pteptll_DPA_ATLAS} we turn to
the transverse-momentum distributions of the electron (left) and the charged-lepton system (right).
\bfi
\includegraphics[angle=0,scale=0.7,bb=60 330 370 770]{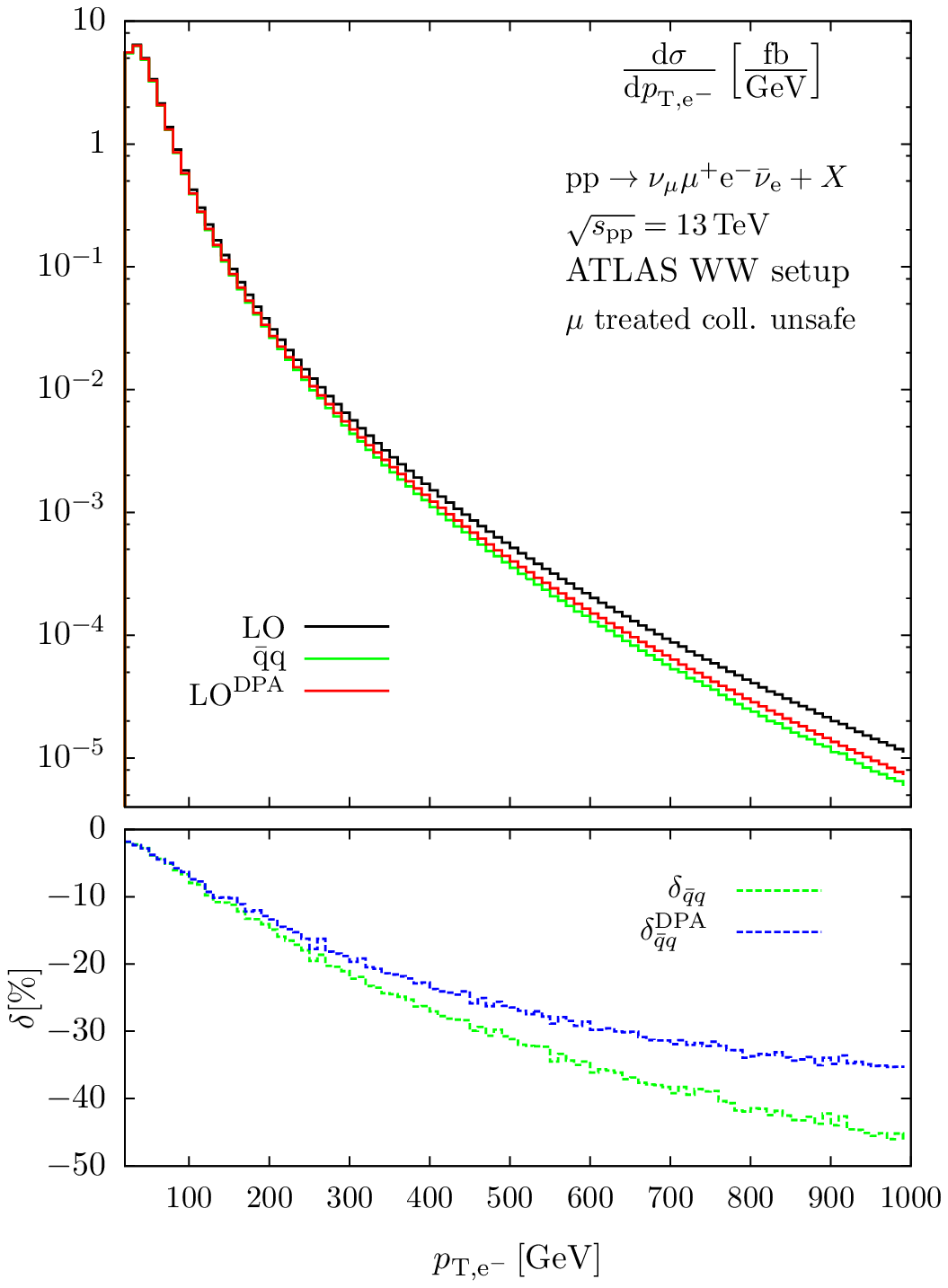}
\includegraphics[angle=0,scale=0.7,bb=60 330 370 770]{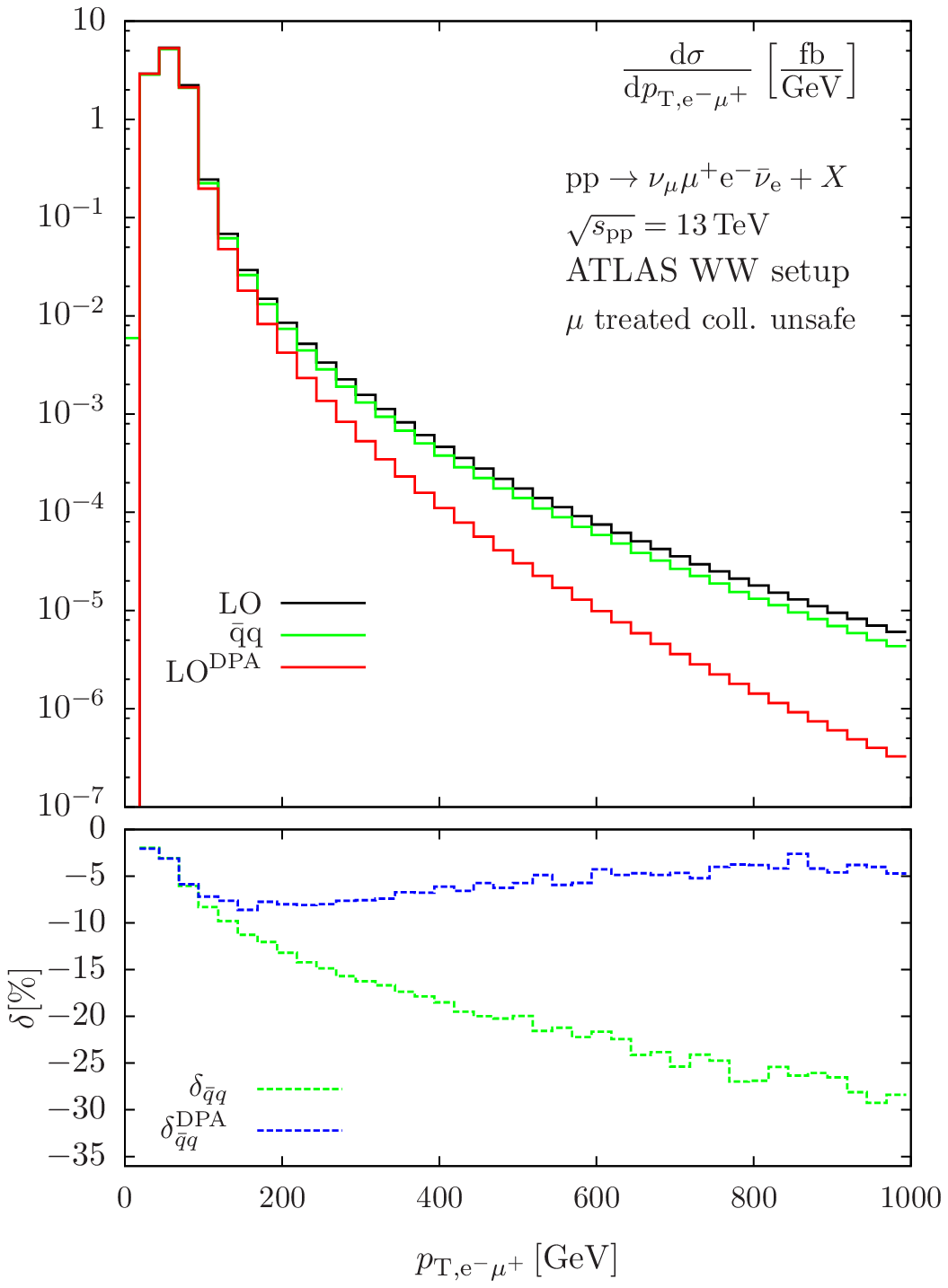}
\caption{Transverse-momentum distributions of the electron (left) and of the charged-lepton system (right) in $\Pp\Pp\to\mvev+X$ in the
ATLAS~WW setup. The lower panels show the relative size of the EW corrections to the $\bar{q}q$ channels
in our default setup compared to the result based on the DPA.
\label{fig:pteptll_DPA_ATLAS}
}
\efi
\bfi
\vspace*{1em}
\centerline{
{\scriptsize
\unitlength=0.5pt%
\begin{feynartspicture}(280,200)(1,1)
\DashLine(100,50)(180,50){5}
\Text(330, 90)[]{\tiny beam axis}
\SetColor{Green}
\LongArrow(140,55)(140,100)
\SetColor{Red}
\LongArrow(140,45)(140,  0)
\SetColor{Black}
\Text(300,165)[l]{\Green{$\vec p_{\rT,\mu^+\nu_\mu\bar\nu_\Pe}$}}
\Text(300,40)[l]{\Red{$\vec p_{\rT,\Pem}$}}
\FADiagram{}
\FAProp(0.,15.)(4.,10.)(0.,){/Straight}{-1}
\FALabel(-.584783,12.8238)[tr]{$\bar q$}
\FAProp(0.,5.)(4.,10.)(0.,){/Straight}{1}
\FALabel(-.584783,7.17617)[br]{$q$}
\FAProp(20.,22.)(16.5,17.)(0.,){/Straight}{-1}
\FALabel(22.5,22.5098)[br]{\Green{$\nu_\mu$}}
\FAProp(21.,20.)(16.5,17.)(0.,){/Straight}{1}
\FALabel(24.,21.7931)[tr]{\Green{$\mu^+$}}
\FAProp(20.,0.)(9.5,10.)(0.,){/Straight}{-1}
\FALabel(23,.29897)[tr]{\Red{$\Pem$}}
\FAProp(20.,16.)(13.,13.5)(0.,){/Straight}{1}
\FALabel(23.,16.6468)[tr]{\Green{$\bar\nu_\Pe$}}
\FAProp(4.,10.)(9.5,10.)(0.,){/Sine}{0}
\FALabel(5.5,9.23)[t]{$\gamma/\PZ$}
\FAProp(16.5,17.)(13.,13.5)(0.,){/Sine}{1}
\FALabel(13.6011,16.1387)[br]{$\PW$}
\FAProp(9.5,10.)(13.,13.5)(0.,){/Straight}{-1}
\FALabel(10.1011,12.6387)[br]{$\Pe$}
\FAVert(4.,10.){0}
\FAVert(16.5,17.){0}
\FAVert(9.5,10.){0}
\FAVert(13.,13.5){0}
\end{feynartspicture}
}
\hspace*{5em}
{\scriptsize
\unitlength=0.5pt%
\begin{feynartspicture}(280,200)(1,1)
\DashLine(100,50)(180,50){5}
\Text(330, 90)[]{\tiny beam axis}
\SetColor{Red}
\LongArrow(140,70)(140,105)
\SetColor{Green}
\LongArrow(140,55)(140, 65)
\LongArrow(140,45)(140,  0)
\SetColor{Black}
\Text(300,170)[l]{\Red{$\vec p_{\rT,\Pem\mu^+}$}}
\Text(300,120)[l]{\Green{$\vec p_{\rT,\bar\nu_\Pe}$}}
\Text(300,40)[l]{\Green{$\vec p_{\rT,\nu_\mu}$}}
\FADiagram{}
\FAProp(0.,15.)(4.,10.)(0.,){/Straight}{-1}
\FALabel(-.584783,12.8238)[tr]{$\bar q$}
\FAProp(0.,5.)(4.,10.)(0.,){/Straight}{1}
\FALabel(-.584783,7.17617)[br]{$q$}
\FAProp(20.,22.)(16.5,17.)(0.,){/Straight}{-1}
\FALabel(22.0,22.5098)[br]{\Red{$\Pem$}}
\FAProp(21.,20.)(16.5,17.)(0.,){/Straight}{1}
\FALabel(23.,21.2931)[tr]{\Green{$\bar\nu_\Pe$}}
\FAProp(20.,0.)(9.5,10.)(0.,){/Straight}{-1}
\FALabel(22.7,-.29897)[tr]{\Green{$\nu_\mu$}}
\FAProp(20.,16.)(13.,13.5)(0.,){/Straight}{1}
\FALabel(22.5,16.9468)[tr]{\Red{$\mu^+$}}
\FAProp(4.,10.)(9.5,10.)(0.,){/Sine}{0}
\FALabel(5.5,9.23)[t]{$\PZ$}
\FAProp(16.5,17.)(13.,13.5)(0.,){/Sine}{1}
\FALabel(13.6011,16.1387)[br]{$\PW$}
\FAProp(9.5,10.)(13.,13.5)(0.,){/Straight}{-1}
\FALabel(10.1011,12.6387)[br]{$\nu_\mu$}
\FAVert(4.,10.){0}
\FAVert(16.5,17.){0}
\FAVert(9.5,10.){0}
\FAVert(13.,13.5){0}
\end{feynartspicture}
}
\hspace*{5em}
}
\caption{Illustration of diagrammatic structures dominating the $p_{\rT,\Pe^-}$
(left) and $p_{\rT,\Pe^-\mu^+}$ (right) distributions shown in 
\reffig{fig:pteptll_DPA_ATLAS} for high transverse momenta.}
\label{fig:singlyresdiags}
\efi
In the $p_{\rT,\Pe^-}$ distribution, the comparison between full and
DPA calculation reveals similar qualitative features as for the
$M_{\Pem\mu^+}$ distribution.  The differences are, however, larger in
size, reaching the $5\%$ ($10\%$) level for transverse momenta
$p_{\rT,\Pe^-}$ of about $500\GeV$ ($1\TeV$).  Again the deterioration
of the DPA can already be seen at LO and attributed to an enhanced
impact of the singly-resonant background diagrams shown in the second
line of \reffig{fig:tree-ddnmen}, which are not included in 
$\sigma^\mr{LO,DPA}_{\bar q q}$.
Schematically the relevant diagrams are kinematically illustrated
on the l.h.s.\ of \reffig{fig:singlyresdiags}.
The enhancement is due to events where one single lepton is recoiling
against the other three 
in the final state. Thus, for large
$p_{\rT,\Pe^-}$ the cross section $\rd\sigma^\mr{LO}_{\bar q q}/\rd
p_{\rT,\Pe^-}$ receives large contributions from events where the
electron is back-to-back to the three other leptons.  For
doubly-resonant diagrams (first line in \reffig{fig:tree-ddnmen}) this
situation is less likely 
for large $p_{\rT,\Pe^-}$, where
the W-decay lepton pairs mostly appear back-to-back as a result of the boost
from the W~rest frames to the laboratory system.
The comparison of $\sigma^\mr{LO}_{\bar q q}$ with
$\sigma^\mr{LO,DPA}_{\bar q q}$ at high $p_{\rT,\Pem}$ shows that
singly-resonant contributions dominate over doubly-resonant parts
already for a $p_{\rT,\Pem}$ of some $100\GeV$.
Kinematically, it is thus easier to produce leptons with high transverse
momenta directly rather than through the decay of W bosons.
The difference between
full and DPA NLO EW corrections is again reproduced quite well
by the last term of our estimate \refeq{eq:DeltaDPA}.

The difference between the full $4f$ and the DPA calculation is pushed
to the extreme in the distribution of the transverse momentum
$p_{\rT,\Pem\mu^+}$ of the charged-lepton system.  Here the
enhancement of background diagrams is due to events where one neutrino
recoils against the two charged leptons and the other neutrino, a
situation that is supported by singly-resonant diagrams
as illustrated on the r.h.s.\ of \reffig{fig:singlyresdiags},
but not by
doubly-resonant graphs, where the two charged leptons tend to
recoil
against each other for high transverse momenta.  Looking at equal
sizes of transverse momenta on the horizontal axes of the two
distributions in \reffig{fig:pteptll_DPA_ATLAS}, the enhancement seems
stronger in the case of the $p_{\rT,\Pem\mu^+}$ distribution, but at
the same time it should be realized that the spectrum on the r.h.s.\ 
drops much faster than the one of $p_{\rT,\Pe^-}$ on the l.h.s.\ for
increasing $p_{\rT}$.  This is due to the fact that
$p_{\rT,\Pem\mu^+}$ contains 
only part of the transverse momentum of the
three-lepton system recoiling against the single neutrino
and that it is very unlikely to produce a large $p_{\rT,\Pem\mu^+}$
via the doubly resonant contributions.
In
conclusion, we can state that transverse-momentum distributions are
reproduced by the DPA only up to some $100\GeV$ owing to the growing
influence of background diagrams that do not show two simultaneously 
resonant W~bosons.  For predictions of such $p_{\rT}$ 
spectra in the TeV
range, the calculation of EW corrections should be based on a full
4-fermion~calculation.

%
%
\section{Conclusions and outlook}
\label{sec:concl}

Electroweak di-boson production processes 
represent a very interesting class of particle reactions at the LHC.
They provide an ideal test-ground for the non-Abelian
self-interactions among the electroweak gauge bosons,
but also form an important class of backgrounds to many
new-physics searches and 
to precision studies of the Higgs boson.
Precise calculations 
for these processes, including
radiative corrections of strong and electroweak interactions, 
are thus phenomenologically very important and have seen great progress
in recent years. On the one hand, QCD predictions are being pushed to
the next-to-next-to-leading-order level and are extended by including
leading corrections beyond fixed orders. On the other hand,
calculations of electroweak corrections become more and more refined
by including decays of the unstable bosons and off-shell effects.

In this paper, we have reported the results from 
the first calculation of
next-to-leading-order electroweak corrections to W-boson pair
production that fully takes into account leptonic W-boson decays and
off-shell effects.  
This calculation is based on the complex-mass
scheme for the treatment of the intermediate electroweak gauge bosons
and provides next-to-leading-order precision over the entire
phase space with resonant and/or non-resonant W~bosons.
Thus, it goes beyond previous calculations which are
restricted to on-shell or nearly resonant W~bosons.

We have discussed the electroweak corrections using different 
event selections, comprising one that is
typical for the study of W-boson pairs as a signal process
and another one that is typical for an analysis of
Higgs-boson decays $\PH\to\PW\PW^*$, where W-boson pair production
represents an irreducible background. 
In particular, we have compared the full off-shell
results to previous results in the so-called
double-pole approximation, which is based on an expansion of the loop amplitudes
about the poles of the W~resonances.
At small and intermediate scales, i.e.\ in particular in angular and
rapidity distributions, the two approaches show the expected
agreement at the level of fractions of a percent, but larger differences appear for invariant-mass and transverse-momentum distributions 
in the TeV range. For transverse-momentum distributions, the differences can
even exceed the 10\% level in the TeV range where ``background diagrams''
with one instead of two resonant W~bosons grow in importance because of 
the suppression of the doubly resonant contributions in these
kinematical regimes.

To fully exploit our calculation in upcoming LHC data analyses, our 
state-of-the-art results on electroweak corrections should be combined
with QCD-corrected cross sections. A simple way to achieve this, 
would be to apply differential reweighting factors for the electroweak
corrections to differential distributions obtained with 
state-of-the-art QCD-based predictions.
Predictions obtained in this way would be 
accurate to the level of very few percent for integrated cross sections and distributions
that are dominated by energy scales up to few $100\GeV$. 
For transverse-momentum and invariant-mass distributions in the TeV range,
the precision will deteriorate and most likely be limited by our
knowledge of QCD corrections, while electroweak corrections are sufficiently well
under control.
%
%
\section*{Acknowledgements}

B.J.\ and L.S.\
are grateful to Julien Baglio for valuable discussions. 
B.B.\ and A.D.\ acknowledge support by the Deutsche
Forschungsgemeinschaft (DFG) under reference number
\mbox{DE~623/2-1}.  
The work of M.B., B.J.\ and L.S.\ is supported in part by the Institutional
Strategy of the University of T\"ubingen (DFG, ZUK~63) and in part by
the German Federal Ministry for Education and Research (BMBF) under
contract number 05H2015.
S.D.\ is supported by the DFG through the Research Training Group 
RTG~2044.  
The work of L.H.\ was
supported by the grants FPA2013-46570-C2-1-P and 2014-SGR-104, and
partially by the Spanish MINECO under the project MDM-2014-0369 of
ICCUB (Unidad de Excelencia ``Mar\'ia de Maeztu'').
Part of this work was performed on the high performance computational resources funded by the Ministry of Science,  Research and the Arts and the Universities of the State of Baden-W\"urttemberg, Germany, within the framework program bwHPC.
%
%

\end{document}